\documentclass[12pt]{article}
\usepackage[T2B]{fontenc}

 \usepackage{amssymb}
\begin{document}

\newcommand{\beqn}{\begin{eqnarray}}
 \newcommand{\eeqn}{\end{eqnarray}}
 \newcommand{\be}{\begin{equation}}
 \newcommand{\ee}{\end{equation}}
 \newcommand{\ba}{\begin{array}}
 \newcommand{\ea}{\end{array}}
 \newcommand{\E}{\mathbb{E}}
\newcommand{\T}{\mathbb{T}}
\newcommand{\R}{\mathbb{R}}
\newcommand{\Z}{\mathbb{Z}}
\newcommand{\N}{\mathbb{N}}
\newcommand{\CC}{\mathbb{C}}

  \newcommand{\de}{\delta}
 \newcommand{\fr}{\frac}
  \newcommand{\pa}{\partial}
 \newcommand{\ov}{\overline}
 \newcommand{\ve}{\varepsilon}
\newcommand{\tr}{{\rm tr~}}
 \newcommand{\ds}{\displaystyle}
\newcommand{\const}{\mathop{\rm const}\nolimits}
\newcommand{\supp}{\mathop{\rm supp}\nolimits}
\newcommand{\sgn}{\mathop{\rm sgn}\nolimits}

\newcommand{\bo}{{\hfill\loota}}
\newcommand{\loota}{\hbox{\enspace{\vrule height 7pt depth 0pt width 7pt}}}

 \renewcommand{\theequation}{\thesection.\arabic{equation}}

 \newtheorem{theorem}{Theorem}[section]
 \newtheorem{definition}[theorem]{Definition}
  \newtheorem{lemma}[theorem]{Lemma}
 \newtheorem{example}[theorem]{Example}
 \newtheorem{remark}[theorem]{Remark}
 \newtheorem{remarks}[theorem]{Remarks}
 \newtheorem{cor}[theorem]{Corollary}
 \newtheorem{pro}[theorem]{Proposition}

\begin{center}
{\Large\bf Behavior for large time of an infinite chain\\~ \\
of harmonic oscillators with  defects}
\vspace{1cm}\\
{\large T.V. Dudnikova}
\medskip\medskip\\
{\it  Keldysh Institute of Applied Mathematics of Russian Academy of Sciences}
\medskip\\
Miusskaya sq. 4, Moscow 125047, Russia  \medskip\\
E-mail:~tdudnikov@mail.ru
\end{center}
\vspace{1cm}

\begin{abstract}
An infinite irregular harmonic chain of particles is considered.
We assume that some particles (``defects'') in the chain have masses
and force constants of interaction
different from the masses and the interaction constants of the other particles.
We study the Cauchy problem for this model.
The main goal is to study the long-time behavior and derive the dispersive bounds for the solutions in the
energy weighted norms.
\medskip

\textbf{\textit{Key words and phases:}}
infinite chain of harmonic oscillators with  defects,
Cauchy problem, Fourier--Laplace transform, Puiseux expansion, dispersive estimates
\medskip


\end{abstract}

 \newpage
\section{Introduction}
We consider a Hamiltonian infinite system of particles having harmonic nearest-neighbor
interactions with the Hamiltonian functional of a form
\be\label{H1}
{\bf H}(u,v)=\ds\frac12
\sum\limits_{n\in\Z}\Big(
\frac{|v(n)|^2}{m_n}+\gamma_n|u(n+1)-u(n)|^2+\mu_n|u(n)|^2\Big),
\ee
where $m_n,\gamma_n>0$, $\mu_n\ge0$. Then, the displacement of the $n$-th particle
from its equilibrium position obeys the following equations:
 \beqn\label{1}
 \left\{\ba{rcl}
 \dot u(n,t)\!\!&=&\!\!\ds\frac{\delta\, {\bf H}}{\delta v}=\frac{v(n,t)}{m_n}
\\
\dot v(n,t)\!\!&=&\!\!\ds-\frac{\delta\, {\bf H}}{\delta u}
=\gamma_n \nabla_Lu(n,t)-\gamma_{n-1} \nabla_Lu(n-1,t)-\mu_n u(n,t)
\ea \right|\,\,\, n\in\Z,\quad t>0.
\eeqn
Here $u(n,t)\in\R$,  $\nabla_L$ denotes the derivative on $\Z=\{0,\pm1,\pm2,\dots\}$,
$$
\nabla_Lu(n)=u(n+1)-u(n),\quad n\in\Z.
$$
 We denote by $\gamma_n$ the force constant of interaction between the nearest neighbors,
by $v(n,t)=m_n\dot u(n,t)$  the moment of the $n$-th particle,
by $\dot u(n,t)$ its velocity.
We fix some $N\ge0$ and
assume that particles located at points  $n\ge N+1$  and $n\le -1$
have the same mass $m_n=m_+>0$ and  $m_n=m_->0$, respectively.
Furthermore,
 particles are affected by the same external harmonic forces with constants
 $\mu_n=\mu_+\ge0$ for $n\ge N+1$ and $\mu_n=\mu_-\ge0$ for $n\le-1$.
 The force constants of interaction are of a form $\gamma_n=\gamma_-$ for
$n\le -1$, $\gamma_n=\gamma_+$ for $n\ge N$.
 At the same time, the particles  (so-called ``defects'')
 located at the points $n=0,1,\dots,N$
 have constants $m_n,\mu_n,\gamma_n$, generally speaking,
 different from $m_\pm,\mu_\pm,\gamma_\pm$.

Therefore, the system (\ref{1}) becomes
\beqn
 m_- \ddot u(n,t)\!\!&=&\!\!\ds
 (\gamma_-\Delta_L-\mu_-) u(n,t), \quad n\le -1,\quad t>0,\label{1-}\\
m_n\ddot u(n,t)\!\!&=&\!\!\gamma_n \nabla_Lu(n,t)-\gamma_{n-1} \nabla_Lu(n-1,t)-\mu_n u(n,t),\,\,
 n=0,1,\dots,N,\,\,\,\,\,\,\label{1x}\\
 m_+ \ddot u(n,t)\!\!&=&\!\!\ds
 (\gamma_+\Delta_L-\mu_+) u(n,t), \quad  n\ge N+1,\quad t>0\label{1+}.
\eeqn
Here $\Delta_L$ denotes the second derivative on $\Z$:
$$
\Delta_L u(n)=u(n+1)-2u(n)+u(n-1)=\nabla_Lu(n)-\nabla_Lu(n-1),\quad n\in\Z.
$$
For  system (\ref{1}), we study the Cauchy problem with the initial data
\beqn\label{1.3}
u(n,0)=u_0(n),\quad v(n,0)=m_n \dot u(n,0)=v_0(n),\quad n\in\Z.
\eeqn
Write
$Y(t)=(u(\cdot,t), v(\cdot,t))$, $Y_0(\cdot)\equiv (Y^0_0(\cdot),Y^1_0(\cdot))=(u_0(\cdot),v_0(\cdot))$.
We assume that the initial data $Y_0$ belong to
the Hilbert space ${\cal H}_{\alpha}$, $\alpha\in\R$, defined below.
 \begin{definition} \label{d1.1'}
 $\ell^2_{\alpha}\equiv\ell^2_{\alpha}(\Z)$,
$\alpha\in\R$, is the  Hilbert space of sequences $u(n)$, $n\in\Z$,
 with  norm
$ \Vert u\Vert^2_{\alpha}
=\sum\limits_{n\in\Z}\langle n\rangle^{2\alpha}|u(n)|^2<\infty$,
$\langle n\rangle:=(1+n^2)^{1/2}$.\\
$ {\cal H}_{\alpha}=\ell^2_{\alpha}\otimes\ell^2_{\alpha}$
is the  Hilbert space of pairs $Y=(u,v)$ of sequences equipped with  norm
 $ \Vert Y\Vert^2_{\alpha}
= \Vert u\Vert^2_{\alpha}+ \Vert v\Vert^2_{\alpha}<\infty$.
 \end{definition}

Our main goal is to find  restrictions on the constants $m_n$, $\gamma_n$, $\mu_n$
 (see conditions ${\bf C}$ and ${\bf C}_0$ in Sec.~\ref{sec2}) under which
 for any initial data $Y_0\in{\cal H}_{\alpha}$ with $\alpha>3/2$,
  the solution $Y(t)$ of  system (\ref{1-})--(\ref{1.3}) obeys the following bound
\be\label{0.3}
\Vert Y(t)\Vert_{-\alpha}\le C(1+|t|)^{-\beta/2}\Vert Y_0\Vert_{\alpha},
\quad t\in\mathbb{R},\quad \alpha>3/2,
\ee
where $\beta=1$ if condition~${\bf C}_0$ holds, and
$\beta=3$ if condition~${\bf C}$ holds.
The last bound is useful for applications to scattering problems.
In particular, we prove that there exists a bounded `wave' operator
$\Omega:{\cal H}_\alpha\to{\cal H}_{-\alpha}$ such that
$$
Y(t)=\Omega(\tilde Y(t))+\delta(t),\quad \mbox{where }\,
\Vert \delta(t)\Vert_{-\alpha}\le C(1+|t|)^{-3/2}\Vert Y_0\Vert_{\alpha}.
$$
Here $\tilde Y(t)\equiv0$ for $n=0,1,\dots,N$
and $\tilde Y(t)$ is a solution of Eqn~(\ref{1-}) for $n\le -1$
and of Eqn~(\ref{1+}) for $n\ge N+1$
with initial data $Y_0$, see  Theorem~\ref{thC}.
\smallskip

Finally, we note that  instead of system (\ref{1-})--(\ref{1+}), it is possible to consider a more general model with
 additional friction terms $-\beta_n\dot u(n,t)$ in Eqn~(\ref{1x}):
$$
m_n\ddot u(n,t)=\gamma_n \nabla_Lu(n,t)-\gamma_{n-1} \nabla_Lu(n-1,t)-\mu_n u(n,t)
-\beta_n\dot u(n,t),\,\,\,    n=0,\dots,N,
$$
where $\beta_n\ge0$. If $\beta_n>0$ for some $n$, then conditions on the constants
could be weakened. This model was studied in \cite{rjmp-2018} in the case
when $N=0$ and $m_\pm=m_n=1$ for all $n$.
In this paper, for simplicity,  we study only the Hamiltonian model~(\ref{1-})--(\ref{1+})
without friction terms.
Furthermore, we consider here the harmonic chain with nearest neighbor interaction.
However, the results can be generalized to a more general case of interaction
between the particles of the chain.

\newpage
\setcounter{equation}{0}
\section{Main Results}\label{sec2}
\subsection{Decomposition of solution}
At first, introduce two initial-boundary value problems with zero boundary condition
\beqn\label{a.1}
\left\{
\ba{ll}
m_\pm\ddot z_\pm(n,t)=(\gamma_\pm\Delta_L-\mu_\pm)z_\pm(n,t),\quad \pm n\ge1,\quad t>0,
\\
z_\pm(0,t)=0,\quad t>0, 
\\
z_\pm(n,0)=u_0(n),\quad m_\pm\dot z_\pm(n,0)=v_0(n),\quad \pm n\ge1. 
\ea\right.
\eeqn
Denote by $ {\cal H}_{\alpha,\pm}=\ell^2_{\alpha,\pm}\otimes\ell^2_{\alpha,\pm}$
 the  Hilbert space of pairs $Y=(u,v)$
of sequences equipped with  norm
 $ \Vert Y\Vert^2_{\alpha,\pm}
= \Vert u\Vert^2_{\alpha,\pm}+ \Vert v\Vert^2_{\alpha,\pm}<\infty$,
where
$\ell^2_{\alpha,\pm}\equiv\ell^2_{\alpha}(\Z_\pm)$,
$\alpha\in\R$, is the  Hilbert space of sequences $u(n)$, $n\in\Z_\pm$,
 with  norm
$ \Vert u\Vert^2_{\alpha,\pm}
=\sum\limits_{\pm n\ge1}\langle n\rangle^{2\alpha}|u(n)|^2<\infty$.
The results concerning the solutions of problem (\ref{a.1})
are stated in \cite{rjmp-2018, preprint18}.
In particular, the following result is proved in \cite{rjmp-2018}.
 \begin{lemma} \label{l2.1} 
Assume that $\alpha\in\R$. Then
(i) for any initial data $Y_0 \in {\cal H}_{\alpha,\pm}$, there exists  a unique solution
$Z_\pm(t)=(z_\pm(\cdot,t),m_\pm\dot z_\pm(\cdot,t))\in C(\R, {\cal H}_{\alpha,\pm})$  to  problem
(\ref{a.1});\\
(ii) the operator  $W_\pm(t):Y_0\mapsto Z_\pm(t)$ is continuous
on ${\cal H}_{\alpha,\pm}$.
Furthermore, the following bound holds,
\be\label{c.2}
\Vert W_\pm(t) Y_0\Vert_{\alpha,\pm}\le C\langle t\rangle^\sigma\Vert Y_0\Vert_{\alpha,\pm}
\ee
with some constants $C=C(\alpha),\sigma=\sigma(\alpha)<\infty$.
\end{lemma}

The proof of Lemma~\ref{l2.1} is based on the following formula for the solutions
of  problem (\ref{a.1}):
\be\label{sol}
Z^{i}_\pm(n,t)=\sum\limits_{\pm k\ge1}
 G^{ij}_{t,\pm}(n,k) Y_{0}^j(k), \quad \pm n\ge1,\quad i=0,1,
\ee
where $Z^{0}_\pm(n,t)\equiv z_\pm(n,t)$, $Z^{1}_\pm(n,t)\equiv m_\pm\dot z_\pm(n,t)$,
$Y^0_0(n)\equiv u_0(n)$, $Y_0^1(n)\equiv v_0(n)$,
the Green function $ G_{t,\pm}(n,k)=( G^{ij}_{t,\pm}(n,k))_{i,j=0}^1$ is a matrix-valued function
of a form
\beqn\label{3.2}
 G_{t,\pm}(n,k):={\cal G}_{t,\pm}(n\!-\!k)-{\cal G}_{t,\pm}(n+k),
\quad
{\cal G}_{t,\pm}(n)\equiv\frac1{2\pi}\int_{\T}
e^{-in \theta} \hat{\cal G}_{t,\pm}(\theta)d\theta,\,\,
\eeqn
$\T\equiv\R/(2\pi\Z)$ denotes torus,
\beqn
\label{hatcalG}
\ba{ll} \hat {\cal G}_{t,\pm}(\theta)
=\left(\hat{\cal G}^{ij}_{t,\pm}(\theta)\right)_{i,j=0,1}
=\left(\ba{ll} \cos(\phi_\pm(\theta)t)&
\sin(\phi_\pm(\theta)t)/\left(m_\pm\phi_\pm(\theta)\right)\\
-m_\pm\phi_\pm(\theta)\sin(\phi_\pm(\theta)t)& \cos(\phi_\pm(\theta)t)
\ea\right)\\
 \phi_\pm(\theta)=\sqrt{\nu^2_\pm(2-2\cos\theta)+\kappa_\pm^2},
 \ea\eeqn
where, by definition,
\be\label{not}
\nu_\pm^2=\gamma_\pm/m_\pm>0,\quad \kappa_\pm^2=\mu_\pm/m_\pm\ge0.
\ee
We see that $z_\pm(0,t)\equiv 0$ for any $t$, since
${\cal G}^{ij}_{t,\pm}(-n)={\cal G}^{ij}_{t,\pm}(n)$.
For the solutions  of  problem~(\ref{a.1}),
the following bound is true.
\begin{theorem}\label{t1} (see \cite[Theorem 2.2]{D16})
Let $Y_0\in{\cal H}_{\alpha,\pm}$ and $\alpha>3/2$.  Then
\beqn \label{ubound}
\Vert W_\pm(t)Y_0\Vert_{-\alpha,\pm}\le
C\langle t\rangle^{-3/2}\Vert Y_0\Vert_{\alpha,\pm},\quad t\in\R.
\eeqn
\end{theorem}

Introduce the boundary-initial value problem in $\Z_N:=\{n\in\Z:n\ge N+1\}$:
\beqn
&&m_+\ddot z_N(n,t)=(\gamma_+\Delta_L-\mu_+)z_N(n,t),\quad  n\in\Z_N,\quad t>0,
\label{N.1}\\
&& z_N(N,t)=0,\quad t>0,
\label{N.2}\\
&&z_N(n,0)=u_0(n),\quad m_+\dot z_N(n,0)=v_0(n),\quad  n\in\Z_N.
\label{N.3}
\eeqn
Denote by $W_N(t)$ the solving operator of this problem,
 \be\label{2.12}
 W_N(t):Y_0\mapsto Z_N(t)=(z_N(\cdot,t),m_+\dot z_N(\cdot,t)).
 \ee
 Then, $\left(W_N(t)Y_0\right)(n)=(W_+(t)\tilde Y_0)(n-N)$
 for $n\in\Z_N$, where  $\tilde Y_0(k):=Y_0(k+N)$ for $k\ge1$,
 $W_+(t)$ is introduced in Lemma~\ref{l2.1}.
Write ${\cal H}_{\alpha,N}$ the Hilbert space of pairs $Y_0=(u_0,v_0)$
with finite norm $\Vert Y_0\Vert_{\alpha,N}<\infty$,
where
 $ \Vert Y_0\Vert^2_{\alpha,N}
= \Vert u_0\Vert^2_{\alpha,N}+ \Vert v_0\Vert^2_{\alpha,N}$,
$ \Vert u\Vert^2_{\alpha,N}
=\sum\limits_{n\in\Z_N}\langle n\rangle^{2\alpha}|u(n)|^2$.
 Hence, for the solutions $z_N(n,t)$
 the following bound holds (cf (\ref{ubound}))
\beqn \label{uboundN}
\Vert W_N(t)Y_0\Vert_{-\alpha,N}\le
C\langle t\rangle^{-3/2}\Vert Y_0\Vert_{\alpha,N},\quad t\in\R,\quad \alpha>3/2,
\eeqn
for any $Y_0\in{\cal H}_{\alpha,N}$.

The following theorem can be proved in a similar way as
\cite[Theorem 2.2]{jmp-2017}.
\begin{theorem}\label{T.A}
(i) Let  $Y_0\in{\cal H}_{\alpha}$, $\alpha\in\R$.
Then the Cauchy problem (\ref{1-})--(\ref{1.3}) has
a unique solution $Y(t)\in C(\R,{\cal H}_{\alpha})$. \\
(ii) The operator
$U(t):Y_0\to Y(t)$ is continuous on ${\cal H}_{\alpha}$. Moreover,
 there exist constants $C,B<\infty$
such that $\Vert U(t)Y_0\Vert_{\alpha}\le Ce^{B|t|}\Vert Y_0\Vert_{\alpha}$.\\
(iii) Let  $Y_0\in{\cal H}_{0}$. Then  the following identity holds:
\be\label{H-1}
{\bf H}(Y(t))={\bf H}(Y_0),\quad t\ge0,
\ee
where ${\bf H}(Y)$ is the Hamiltonian  defined in (\ref{H1}).
\end{theorem}

Below we assume that $\alpha>3/2$.

We represent  the solution of  problem (\ref{1-})--(\ref{1.3})
in the following form
\be\label{2.1}
u(n,t)=z(n,t)+r(n,t),\quad n\in\Z, \quad t>0,
\ee
where, by definition,
\be\label{2.3}
z(n,t)=\left\{\ba{ll}
z_-(n,t)&\mbox{if }\,\, n\le-1,\\
0&\mbox{if }\,\, n=0,\dots,N,\\
z_N(n,t)&\mbox{if }\,\, n\ge N+1.
\ea\right.
\ee
 Therefore, $r(n,t)$  is a solution of the following problem
\beqn
m_-\ddot r(n,t)&=&(\gamma_-\Delta_L-\mu_-) r(n,t),\quad n\le-1,\quad t>0,\label{b.1}\\
m_n\ddot r(n,t)&=&\gamma_n \nabla_L r(n,t)- \gamma_{n-1} \nabla_L r(n-1,t) -\mu_n\, r(n,t) \nonumber\\
&&+
\delta_{n0}\,\gamma_-\, z_-(-1,t)+\delta_{nN}\,\gamma_+\,z_N(N+1,t),\,\, n=0,\dots,N,\,\, t>0,\label{b.2}\\
m_+\ddot r(n,t)&=&(\gamma_+\Delta_L-\mu_+) r(n,t),\quad n\ge N+1,\quad t>0,\label{b.1+}\\
r(n,0)&=&0,\quad \dot r(n,0)=0
\quad \mbox{for }\,\,n\le-1\,\, \mbox{and }\,\, n\ge N+1, \label{b.3}\\
r(n,0)&=&u_{0}(n),\quad m_n\dot r(n,0)=v_{0}(n)
\quad \mbox{for }\,\,n=0,\dots N. \label{b.4}
\eeqn
Here $\delta_{ij}$ denotes the Kronecker symbol.


\subsection{The problem in the Fourier--Laplace transform}

To construct the solutions of problem (\ref{b.1})--(\ref{b.4})
we use the Fourier--Laplace transform.
\begin{definition}
Let $|r(t)|\le Ce^{B t}$.
The Fourier--Laplace transform of $r(t)$ is given by the formula
$$
\tilde r(\omega)=\int_{0}^{+\infty}
e^{i\omega t}r(t)\,dt,\quad \Im\omega>B.
$$
\end{definition}
The Gronwall inequality
implies standard a priori estimates for the solutions $r(n,t)$, $n\in\Z$.
In particular, there exist constants $C,B<\infty$ such that
$$
\sum\limits_{n\in\Z}(|r(n,t)|^2+|\dot r(n,t)|^2)\le C e^{Bt}
\quad\mbox{ as }\,\, t\to+\infty.
$$
At first, we study the solutions $r(n,t)$ for $n\notin\{0,\dots,N\}$.
The Fourier--Laplace  transform of $r(n,t)$ with respect to $t$-variable,
$r(n,t)\to\tilde r(n,\omega)$, exists at least
for $\Im \omega>B$ and satisfies the following equation
\beqn\label{b.1'}
(-\nu^2_\pm\Delta_L+\kappa_\pm^2-\omega^2)\tilde r(n,\omega)=0
\,\,\,\, \mbox{for }\,n\le-1 \,\,\mbox{and }\, n\ge N+1,\,\, \Im\omega>B,
\eeqn
where $\nu^2_\pm$ and  $\kappa_\pm^2$  are defined in (\ref{not}).
 Now  we construct the solution of (\ref{b.1'}).
Note that the Fourier transform of the lattice operator $-\nu_\pm^2\Delta_L+\kappa_\pm^2$
is the operator of multiplication by the function
$\phi^2_\pm(\theta)=\nu^2_\pm(2-2\cos\theta)+\kappa_\pm^2$.
Thus, $-\nu^2_\pm\Delta_L+\kappa_\pm^2$ is a self-adjoint operator
and its spectrum is absolutely continuous and coincides with the range of $\phi^2_\pm(\theta)$,
i.e., with the segment $[\kappa_\pm^2,a^2_\pm]$, $a^2_\pm:=\kappa_\pm^2+4\nu_\pm^2$.

We introduce a {\em critical set} $\Lambda$:
\be\label{Lambdapm}
\Lambda=\Lambda_+\cup\Lambda_-,\quad
\Lambda_\pm:=[-a_\pm,-\kappa_\pm] \cup[\kappa_\pm,a_\pm],\quad a_\pm=\sqrt{\kappa_\pm^2+4\nu_\pm^2},
\ee
and denote
$\Lambda^0=\Lambda^0_-\cup\Lambda^0_+$, where $\Lambda^0_\pm:=\{-a_\pm,-\kappa_\pm,\kappa_\pm,a_\pm\}$.
\begin{lemma}\label{theta} (see \cite[Lemma 2.1]{KKK})
For given $\omega\in \mathbb{C}\setminus \Lambda_\pm$,
the equation
\be\label{32}
\nu^2_\pm(2-2\cos\theta)+\kappa^2_\pm=\omega^2
\ee
has the unique solution  $\theta_\pm(\omega)$
in the domain $\{\theta\!\in\!\mathbb{C}\!:\Im\theta\!>\!0,\,\Re\theta\in\!(-\pi,\pi]\}$.
Moreover, $\theta_+(\omega)$ ($\theta_-(\omega)$)
is an analytic function in $\mathbb{C}\setminus \Lambda_+$
(in $\mathbb{C}\setminus \Lambda_-$, respectively).
\end{lemma}

Since we seek the solution  $r(\cdot,t)\in \ell^2_{\alpha}$
with some $\alpha$, $\tilde r(n,\omega)$ has a form
$$
\tilde r(n,\omega)=\left\{
\ba{ll}
e^{-i\theta_-(\omega)n}\,\tilde r(0,\omega) & \mbox{if }\,n\le-1,\\
e^{i\theta_+(\omega)(n-N)}\,\tilde r(N,\omega) & \mbox{if }\,
 n\ge N+1.
\ea\right.
$$
Write $\tilde \Gamma^\pm_n(\omega)=e^{\pm i\theta_\pm(\omega)n}$ for $\pm n\ge1$.
Applying the inverse Fourier--Laplace transform with respect to $\omega$-variable,
we write the solution  of (\ref{b.1}), (\ref{b.3}) in the form
\be\label{qxt}
r(n,t)=\int_{0}^t \Gamma^-_n(t-s)r(0,s)\,ds\quad
\mbox{for }\,\,n\le-1,\quad  t>0,
\ee
and the solution  of (\ref{b.1+}), (\ref{b.3}) in the form
\be\label{qxt+}
r(n,t)=\int_{0}^t \Gamma^+_{n-N}(t-s)r(N,s)\,ds\quad
\mbox{for }\,\,n\ge N+1,\quad  t>0,
\ee
where
\beqn\label{K(x,t)}
\Gamma_n^\pm(t):=\frac1{2\pi}\int\limits_{-\infty+ic}^{+\infty+ic}
e^{-i\omega t}\,\tilde \Gamma^\pm_n(\omega)\,d\omega
\quad \mbox{with some }\,c>0,\quad \pm n\ge1,\quad t>0.
\eeqn

\begin{theorem}\label{l2.15} (see \cite[Theorem 3.3]{jmp-2017})
For any $\alpha>3/2$, the following bounds hold,
 \be\label{boundK}
 \Vert \Gamma^+_n(t)\Vert_{-\alpha,+}\le C\langle t\rangle^{-3/2},\quad
 \Vert \Gamma^-_n(t)\Vert_{-\alpha,-}\le C\langle t\rangle^{-3/2}, \quad t>0.
 \ee
In particular,
\be\label{3.7'}
|\Gamma^+_1(t)|\le C(1+t)^{-3/2},\quad |\Gamma^-_{-1}(t)|\le C(1+t)^{-3/2},\quad t>0.
\ee
\end{theorem}

To study the solution $r(n,t)$ of Eqn~(\ref{b.2}) for $n=0,\dots,N$
we first consider the solutions of
  the corresponding homogeneous equation
\be\label{qt}
m_n\ddot r(n,t)=\gamma_n \nabla_L r(n,t)- \gamma_{n-1} \nabla_L r(n-1,t) -\mu_n\, r(n,t),
\quad t>0,
\ee
where $n=0,\dots,N$,
 $\gamma_{-1}\equiv \gamma_{-}$, $\gamma_{N}\equiv \gamma_{+}$,
 with the initial data (\ref{b.4}).
Applying the Fourier--Laplace transform
to the solutions of problem~(\ref{qt}), (\ref{b.4}), we obtain
\be\label{2.26}
\tilde {\bf r}(\omega)= \tilde {\cal N}(\omega)
\left({\bf v}_0-i\omega{\bf u}_1\right)
 \quad \mbox{for }\,\,\Im\omega>B,
\ee
where $\tilde {\bf r}(\omega)$ denotes a column
$\tilde {\bf r}(\omega)=(\tilde r(0,\omega),\dots,\tilde r(N,\omega))^T$
and
\be\label{2.27}
{\bf u}_1:=(m_0u_0(0),\dots,m_Nu_0(N))^T,\quad
{\bf v}_0:=(v_0(0),\dots,v_0(N))^T,
\ee
 $\tilde {\cal N}(\omega)=(\tilde D(\omega))^{-1}$.
 If $N\ge1$, then  $\tilde D(\omega)=(\tilde D_{kn}(\omega))_{k,n=0}^N$, $\omega\in\mathbb{C}_+$,
 is a tridiagonal symmetric  matrix  with entries of a form
\beqn\label{3.21}
\ba{llll}
\tilde D_{00}(\omega)&=&\mu_0-m_0\omega^2+\gamma_-(1-e^{i\theta_-(\omega)})+\gamma_0,&\\
\tilde D_{nn}(\omega)&=&\mu_n-m_n\omega^2+\gamma_n+\gamma_{n-1},& n=1,\dots,N-1,\\
\tilde D_{NN}(\omega)&=&\mu_N-m_N\omega^2+\gamma_+(1-e^{i\theta_+(\omega)})+\gamma_{N-1}&\\
\tilde D_{nn+1}(\omega)&=&\tilde D_{n+1n}(\omega)=-\gamma_n,& n=0,\dots,N-1,\\
\tilde D_{kn}(\omega)&=&0 &\mbox{for }\,\,|k-n|\ge2.
\ea
\eeqn
If $N=0$, then
\be\label{D-N0}
\tilde D(\omega)=\mu_0-m_0\omega^2+\sum_\pm\gamma_\pm(1-e^{i\theta_\pm(\omega)}),\quad \omega\in\CC_+.
\ee

\subsection{Conditions on the constants}
The properties of the function $\tilde D(\omega)$, $\omega\in\CC$,
 play a key role in the proof of the bound~(\ref{0.3}).
 This function  is studied in Appendices~A and B.
In particular, we check that $\textrm{det }\tilde D(\omega)\not=0$
for any $\omega\in \mathbb{C}_\pm=\{\omega\in\mathbb{C}:\pm\Im\omega>0\}$.
Also, we prove  that
 $\textrm{det }\tilde D(\omega\pm i0)\not=0$  for any $\omega\in\Lambda\setminus\Lambda^0$,
where $\tilde D(\omega\pm i0):=\lim\limits_{\ve\to+0}\tilde D(\omega\pm i\ve)$.
 However, for some constants $m_\pm, \gamma_\pm,\mu_\pm,m_n,\gamma_n,\mu_n$,
 $\textrm{det }\tilde D(\omega)=0$ at some point
 $\omega\in(\R\setminus\Lambda)\cup\Lambda^0$.
 Therefore, to obtain the bound~(\ref{0.3}) we have to find and
eliminate such values of  the constants.
We divide all  values of the constants  into three groups.
The first group ({\em condition}~{\bf C})
includes all values for which $\textrm{det }\tilde D(\omega)\not=0$ for any $\omega\in\R$.
The second group includes values
under which $\textrm{det }\tilde D(\omega)\not=0$ for $\omega\in\R\setminus\Lambda^0$,
 there exists a point $\omega_0\in\Lambda^0$ such that
$\textrm{det }\tilde D(\omega_0)=0$
and $\textrm{det }\tilde D(0)\not=0$ if $0\in\Lambda^0$
(we call these restrictions  by {\em condition}~${\bf C}_0$).
The remaining values of the constants we call the {\em resonance cases}.
For example, the case of the homogeneous chain without external forces
(i.e., when  $m_\pm=m_n$, $\mu_\pm=\mu_n=0$ for $n=0,\dots,N$
 and $\gamma_\pm=\gamma_n$ for $n=0,\dots,N-1$) is a resonance case,
 since in this case $0\in\Lambda^0$ and $\textrm{det }\tilde D(0)=0$.
Thus,  we impose the following conditions ${\bf C}$ or ${\bf C}_0$ on the system.
\smallskip\\
{\bf Condition} ${\bf C}$: $\textrm{det }\tilde D(\omega)\not=0$ for $\omega\in(\R\setminus \Lambda)\cup\Lambda^0$.
\medskip\\
{\bf Condition} ${\bf C}_0$: The following three restrictions hold.
\begin{itemize}
  \item [1)]
   $\textrm{det }\tilde D(\omega)\not=0$ for $\omega\in\R\setminus \Lambda$.
 \item [2)]
 There exists some $\omega_0\in\Lambda^0\setminus \{0\}$ such that
 $\textrm{det }\tilde D(\omega_0)=0$.
 \item [3)] If $\mu_-=\mu_+=0$,  then  $\textrm{det }\tilde D(0)\not=0$.

\end{itemize}

Our main objective is to derive conditions ${\bf C}$ and ${\bf C}_0$
in the terms of the  restrictions on the constants.
 For example,  the third restriction in condition~${\bf C}_0$
is equivalent to the condition that
if  $\mu_\pm=0$,  then
 $\mu_n\not=0$ for some $n\in\{0,\dots,N\}$.

 To state conditions ${\bf C}$ and ${\bf C}_0$ in the case $N=0$,
we introduce the functions $K_0(\omega)$ and $K_\pm(\omega)$ by the rule
\be\label{K0}
K_0(\omega):=\bar\kappa^2-\frac12\sqrt{\kappa_+^2-\omega^2}\sqrt{a_+^2-\omega^2},
\quad \omega\in\R:\,\,\, |\omega|\le\kappa_+;
\ee
where $\bar\kappa:=\left((\kappa_-^2+\kappa_+^2)/2\right)^{1/2}$,
$\kappa_\pm=(\mu_\pm/m_\pm)^{1/2}$, $\nu_\pm=(\gamma_\pm/m_\pm)^{1/2}$,
 $a_\pm=(\kappa^2_\pm+4\nu^2_\pm)^{1/2}$,
\be \label{Kpm}
 K_\pm(\omega):=\bar\kappa^2+\frac12\sqrt{\omega^2-\kappa_\pm^2}\sqrt{\omega^2-a_\pm^2},
 \quad \omega\in\R:\,\,\,|\omega|\ge a_\pm.
     \ee
\begin{theorem}\label{tB}
Let $N=0$. Then  condition~{\bf C} is  the following restrictions.
\beqn\nonumber
\ba{llllll}
&&\mu_0\not=0\,\,\,\mbox{if }\,\, \mu_\pm=0;\\
&&\mu_0<\kappa^2_-(m_0-m_+)+4\nu_-^2\left(m_0-\frac{m_-+m_+}{2}\right)+m_+K_+(a_-)\,\,\,\mbox{if }\,\,\,a_-\ge a_+;\\
&&\mu_0<\kappa_+^2(m_0-m_-)+4\nu_+^2\left(m_0-\frac{m_-+m_+}{2}\right)
+m_-K_-(a_+)\,\,\,\mbox{if }\,\,\,a_+\ge a_-;\\
&&\mu_0>\kappa_-^2(m_0-m_-)+m_+ K_0(\kappa_-)\,\,\,\mbox{if }\,\,\,\kappa_-\not=0;\\
&&
\mu_0>\kappa_+^2(m_0-m_-)+m_- K_-(\kappa_+)\,\,\,\,\mbox{or }\,\,\\
&&
\mu_0<\kappa_-^2(m_0-m_+)+ 4\nu_-^2(m_0-\frac{m_-+m_+}{2})+m_+K_0(a_-)
\,\,\,\mbox{if }\,\,a_-<\kappa_+.
\ea
\eeqn
 Condition~${\bf C}_0$ is the following restrictions:
the inequalities (\ref{4.1})--(\ref{4.4}) (see Appendix~A) hold
and one of the following conditions is fulfilled.
\begin{description}
\item[(i)]
  $a_->a_+$,
  $\mu_0=\kappa^2_-(m_0-m_+)+4\nu_-^2\left(m_0-\frac{m_-+m_+}{2}\right)+m_+K_+(a_-)$

\item[(ii)]
   $a_-<a_+$,  $\mu_0=\kappa_+^2(m_0-m_-)+4\nu_+^2\left(m_0-\frac{m_-+m_+}{2}\right)
+m_-K_-(a_+)$

\item[(iii)]
   $a_+=a_-$, $(\kappa_-,\kappa_+)\not=(0,0)$, $\mu_0=m_0a_-^2- 2\gamma_--2\gamma_+$

\item[(iv)]
   $\kappa_-\not=0$,  $\mu_0=\kappa_-^2(m_0-m_+)+m_+K_0(\kappa_-)$.

In particular, if $\kappa_-=\kappa_+\not=0$, then $\mu_0=\kappa_-^2m_0=\mu_-m_0/m_-$

\item[(v)]
 $a_-\le\kappa_+$,  $\mu_0=\kappa_+^2(m_0-m_+)+m_-K_-(\kappa_+)$

\item[(vi)]
 $a_-<\kappa_+$,
 $\mu_0=\kappa_-^2(m_0-m_+)+4\nu_-^2\left(m_0-\frac{m_-+m_+}{2}\right)+m_+K_0(a_-)$.

\end{description}
\end{theorem}

Theorem~\ref{tB} is proved in Appendix~A.
Note that condition ${\bf C}$ excludes the case when $m_\pm=m_0$, $\gamma_-=\gamma_+$, $\mu_-=\mu_+$.
However, the case when $m_\pm=m_0$, $\gamma_-=\gamma_+$,
$\mu_-=\mu_+\not=0$ is included in condition~${\bf C}_0$.
\begin{remark}\label{r2.8}
Let us consider a particular case of the chain, which we call ({\bf P1})~case,
when  $N=0$ and the oscillators of the chain are identical, excluding defects, i.e.,
\be\label{2.30}
m_-=m_+=: m,\quad \gamma_-=\gamma_+=:\gamma, \quad\mu_-=\mu_+=:\mu.
\ee
Set $\nu=\sqrt{\gamma/m}$, $\nu_0=\sqrt{\gamma/m_0}$,
 $\kappa=\sqrt{\mu/m}$, $\kappa_0=\sqrt{\mu_0/m_0}$,
 $a\equiv a_\pm=\sqrt{(\mu+4\gamma)/m}$, $a_0=\sqrt{(\mu_0+4\gamma)/m_0}$.
In this case,
conditions ${\bf C}$ and ${\bf C}_0$ are simplified as follows
\smallskip
 \begin{itemize}
  \item [${\bf C}$]   $\kappa_0>\kappa$ and $a_0<a$.

\item [${\bf C}_0$] One of the following restrictions holds.

(i) $\kappa_0=\kappa\not=0$ and $\nu_0=\nu$ (and hence, $a=a_0$);

(ii) $\kappa_0=\kappa\not=0$ and $a_0<a$;

(iii) $\kappa_0>\kappa$ and $a_0=a$.
 \end{itemize}

In particular, it follows from conditions~{\bf C} and $\mathbf{C}_0$ that
$\kappa_0\not=0$, $\kappa_0\ge\kappa$ and $\nu_0\le \nu$.
In this particular case,
condition~{\bf C} and $\mathbf{C}_0$  can be rewritten in the terms $m_0,m,\gamma,\mu_0,\mu$ as follows
\begin{itemize}
  \item [${\bf C}$]

  $\ds\frac{\mu_0}{m_0}-\frac{\mu}{m}>0$ and
  $\ds\frac{\mu_0}{m_0}-\frac{\mu}{m}+4\gamma\left(\frac{1}{m_0}-\frac{1}{m}\right)<0$.
\item [${\bf C}_0$] One of the following restrictions hold.

(i) $\mu_0=\mu\not=0$  and $m_0=m$;

(ii) $\mu_0\not=0$, $\mu\not=0$,
 $\ds\frac{\mu_0}{m_0}=\frac{\mu}{m}$ and $m_0>m$;

(iii) $\mu_0\not=0$, $\ds\frac{\mu_0}{m_0}>\frac{\mu}{m}$ and
$\ds\frac{\mu_0}{m_0}-\frac{\mu}{m}+4\gamma\left(\frac{1}{m_0}-\frac{1}{m}\right)=0$.
\end{itemize}
Condition~{\bf C} implies that $m_0>m$ and $\mu_0>\mu\ge0$.
Condition~${\bf C}_0$ implies that
$m_0\ge m$ and $\mu_0\ge \mu$, $\mu_0>0$.
\end{remark}

We see that conditions~${\bf C}$ and ${\bf C}_0$ are tedious even for $N=0$.
Therefore, in the case $N\ge1$,  we assume, in addition,
 that  the oscillators in the chain are identical
except  the defects, i.e., (\ref{2.30}) holds.
In this case, we derive conditions ${\bf C}$ and ${\bf C}_0$
as  restrictions on the constants $m,\gamma,\mu,m_n,\gamma_n,\mu_n$
in Theorem~\ref{tB1}.

Write $a=\sqrt{(\mu+4\gamma)/m}$, $\kappa=\sqrt{\mu/m}$.
Let $\tilde D(\kappa)$ ($\tilde D(a)$) denote the matrix $\tilde D(\omega)$ with
$\omega=\kappa$ ($\omega=a$, resp.), where in the entries we put $e^{i\theta(\kappa)}:=1$ ($e^{i\theta(a)}:=-1$, resp.).
Denote by $\alpha_n(\omega)$, $n=0,1,\dots,N$, the principal (corner) minors of the matrix $\tilde D(\omega)$, see formula~(\ref{alphai}) below.

\begin{theorem}\label{tB1}
Let $N\ge1$ and (\ref{2.30}) hold. Then  conditions~${\bf C}$ and ${\bf C}_0$ are  the following restrictions.
\begin{itemize}
\item [${\bf C}$]
  1) If  $\mu=0$, then  $\mu_n\not=0$ for some $n\in\{0,\dots,N\}$.

2) The matrix $\tilde D(a)$ is negative--definite, $\tilde D(a)<0$.

3) If $\mu\not=0$, then the matrix $\tilde D(\kappa)$ is positive--definite,
$\tilde D(\kappa)>0$.
\end{itemize}
\begin{itemize}
\item [${\bf C}_0$]
   If  $\mu=0$, then   $\mu_n\not=0$ for some $n\in\{0,\dots,N\}$.
    Moreover, the principal minors of the matrix $\tilde D(a)$
    have the following property:
    $\alpha_n(a)(-1)^n<0$ for $n=0,1,\dots,N-1$, and
    $\alpha_N(a)\equiv\textrm{det }\tilde D(a)=0$.

 If $\mu\not=0$, then either (i) $\alpha_n(a)(-1)^n<0$ for $n=0,1,\dots,N-1$,
    $\alpha_N(a)=0$,  and $\tilde D(\kappa)>0$
  or
  (ii) $\alpha_n(\kappa)>0$ for $n=0,1,\dots,N-1$,
    $\alpha_N(\kappa)=0$, and $\tilde D(a)<0$.
\end{itemize}
\end{theorem}

Theorem~\ref{tB1} is proved in Appendix~B.
The proof of this theorem is based on the properties of the tridiagonal matrix $\tilde D(\omega)$.

\begin{remark}
It follows from conditions ${\bf C}$ and ${\bf C}_0$
that $\tilde D_{nn}(a)\le 0$ and
$\tilde D_{nn}(\kappa)\ge0$ for all $n$.
In addition, in the case $N\ge1$,
it follows that the masses of the first and last defects must be greater than
half the mass of the ``non-defective'' particles,
$m_0>m/2$ and  $m_N> m/2$, and also, $\mu_0>\mu/2-\gamma_0$ and
 $\mu_N>\mu/2-\gamma_{N-1}$.
For details, see Appendix~B.
\end{remark}

\subsection{Dispersive bounds}

Define an operator $\overline{W}(t)$, $t\in\R$, on the space ${\cal H}_\alpha$ by the rule
\be\label{2.21}
(\overline{W}(t)Y_0)(n)=
\left\{\ba{ll}
\left(W_-(t)Y_0\right)(n)
&\mbox{for }\,\, n\le-1,\\
0&\mbox{for }\,\,  n=0,\dots N,\\
\left(W_N(t)Y_0\right)(n)
&\mbox{for }\,\,  n\ge N+1,
\ea\right.
\ee
where the operators $W_-(t)$ and $W_N(t)$ are introduced in Lemma~\ref{l2.1}
and in (\ref{2.12}), respectively.
The main result is the following theorem.
\begin{theorem}\label{thC}
 Let $Y_0\in{\cal H}_{\alpha}$, $\alpha>3/2$, and conditions {\bf C} or ${\bf C}_0$ hold.
 Then  the following assertions are fulfilled.
 \begin{itemize}
\item[(i)] There exists a bounded operator $\Omega:{\cal H}_{\alpha}\to{\cal H}_{-\alpha}$
such that
\be\label{2.38}
U(t)Y_0=\Omega(\overline{W}(t)Y_0)+\delta(t),
\quad \mbox{where }\,\,
\Vert \delta(t)\Vert_{-\alpha}\le C\langle t\rangle^{-\beta/2}\Vert Y_0\Vert_{\alpha},
\ee
 $\beta=3$ if condition {\bf C} holds and
$\beta=1$ if condition ${\bf C}_0$ holds.

\item[(ii)]
 $\Vert \Omega(\overline{W}(t)Y_0)\Vert_{-\alpha}\le C\langle t\rangle^{-\beta/2}\Vert Y_0\Vert_{\alpha}$.
 Hence,
  \begin{equation}\label{2.31'}
  \Vert Y(t)\Vert_{-\alpha}\le C\langle t\rangle^{-\beta/2}\Vert Y_0\Vert_{\alpha},
\end{equation}
 where
  $\beta=3$ if condition {\bf C} holds and
$\beta=1$ if condition ${\bf C}_0$ holds.
\end{itemize}
\end{theorem}

Theorem~\ref{thC} can be proved using the technique from \cite{rjmp-2018}.
We outline the  main  steps of the proof.

{\em Step (1)}:
We first apply the Fourier--Laplace transform to the solutions
of problem~(\ref{qt}), (\ref{b.4}) and obtain Eqn~(\ref{2.26}).
Then, we study the behavior of $\tilde {\cal N}(\omega)$
for different values of $\omega\in\CC$:
for $\omega\in\CC_\pm$, $\omega\in\Lambda\setminus\Lambda^0$,
$\omega\in\R\setminus\Lambda$ and for $\omega\in\Lambda^0$,
see Appendices~A and B.
This allows us to
 prove the  bound for the matrix ${\cal N}(t)=\left({\cal N}_{nk}(t)\right)_{n,k=0}^N$, where
\be\label{N}
{\cal N}_{nk}(t)=\frac1{2\pi}\int\limits_{-\infty+ic}
^{+\infty+ic}e^{-i\omega t} \tilde {\cal N}_{nk}(\omega)\,d\omega
\quad \mbox{with some }\,c>0, \quad t>0,
\ee
 $\tilde {\cal N}_{nk}(\omega)$ are entries of the matrix $\tilde {\cal N}(\omega)=(\tilde D(\omega))^{-1}$,
 $n,k=0,\dots,N$.
\begin{theorem}\label{l3.1}
Let  condition {\bf C} or ${\bf C}_0$ hold. Then
\be\label{NN}
|{\cal N}^{(j)}_{nk}(t)|\le C(1+t)^{-\beta/2},\quad t\ge0,\quad j=0,1,2,
\ee
where ${\cal N}^{(j)}_{nk}(t)\equiv\frac{d^j}{dt^j}{\cal N}_{nk}(t)$,
$\beta=3$ if condition {\bf C} holds and
$\beta=1$ if condition ${\bf C}_0$ holds.
\end{theorem}

We prove this theorem in Appendix~A for $N=0$ and in Appendix~B for $N\ge1$.
\medskip

{\em Step (2)}:
Applying the inverse Fourier--Laplace transform to Eqn~(\ref{2.26}),
we write the solution ${\bf r}(t)=(r(0,t),\dots, r(N,t))$
of problem~(\ref{qt}), (\ref{b.4}) in a form
$$
{\bf r}(t)=\dot{\cal N}(t) {\bf u}_1+ {\cal N}(t) {\bf v}_0,
\quad t>0,
$$
where ${\bf u}_1$ and ${\bf v}_0$ are defined in (\ref{2.27}).
For $t=0$ we put ${\cal N}_{nk}(0)=0$ and
 $\dot {\cal N}_{nk}(0)=\delta_{nk}/m_k$ for $n,k=0,\dots,N$.
Using the variation constants formula, we obtain the following representation
for the solutions ${\bf r}(t)=(r(0,t),\dots,r(N,t))$
of Eqs~(\ref{b.2}) with the initial data (\ref{b.4}):
\be\label{2.34}
{\bf r}(t)=\dot {\cal N}(t){\bf u}_{1}+{\cal N}(t){\bf v}_0
+\int_0^t
{\cal N}(t-\tau) {\cal F}(\tau)\,d\tau.
\ee
where
${\cal F}(\tau)=({\cal F}_n(\tau))_{n=0}^{N}$ is a column with entries
${\cal F}_0(\tau)=\gamma_- z_-(-1,\tau)$,
${\cal F}_n(\tau)=0$ for $n=1,\dots,N-1$,
${\cal F}_N(\tau)=\gamma_+z_N(N+1,\tau)$.
 Applying the bounds (\ref{ubound}) and (\ref{uboundN}) to ${\cal F}(\tau)$,
using (\ref{2.34}) and the bound~(\ref{NN}), we obtain the bound for ${\bf r}(t)$:
\be\label{2.35}
\sup_{n=0,1,\dots, N}
\left(|r(n,t)|+|\dot r(n,t)|\right)\le C(1+t)^{-\beta/2}\Vert Y_0\Vert_{\alpha},\quad t\ge0,
\quad \alpha>3/2,
\ee
where  $\beta$ is introduced in Theorem~\ref{l3.1}.
\medskip

{\em Step (3)}: Using formulas (\ref{qxt}), (\ref{qxt+}), (\ref{boundK}) and (\ref{2.35}),
we obtain the following estimate
\be\label{2.5}
\Vert \left(r(\cdot,t),\dot r(\cdot,t)\right)\Vert_{-\alpha}
\le C(1+t)^{-\beta/2}\Vert Y_0\Vert_{\alpha},\quad t\ge0.
\ee
Finally, the bound~(\ref{2.31'}) follows from the decomposition~(\ref{2.1}),
Theorem~\ref{t1} and the bound (\ref{2.5}).
The construction of the operator $\Omega$ and the proof of Theorem~\ref{thC} are
 given in Sec.~\ref{sec4}.
\begin{remark}
Let conditions~${\bf C}$ and ${\bf C}_0$
 be not hold. Then $\tilde {\cal N}_{nk}(\omega)$ have either
 the simple pole at zero or poles
 at points $\omega=\pm\omega_*$, where a point $\omega_*\in\R\setminus\Lambda$ such that
 $\textrm{det }\tilde D(\omega_*)=0$.
 Let the initial data $Y_0\not\equiv0$
 and $Y_0(n)\equiv0$ for $n\not\in\{0,\dots,N\}$.
Therefore, $z(n,t)\equiv0$ for any $n$
and there exist solutions ${\bf r}(t)$ of problem~(\ref{qt}),
(\ref{b.4}) which don't satisfy the bound~(\ref{2.35}).
Hence, there are solutions ${\bf r}(t)$ of (\ref{2.34})
which don't satisfy the bound~(\ref{2.35}).
Thus,
 there exist solutions $u(\cdot,t)$  of  problem~(\ref{1-})--(\ref{1.3})
which do not satisfy the bound~(\ref{2.31'}). For details, see Sections~\ref{Sec3.6}
and \ref{Sec4.6} below.
\end{remark}

\setcounter{equation}{0}
\section{Wave operator}\label{sec4}
Denote by $W'_-(t)$ the operator adjoint to $W_-(t)$, $t\in\R$,
$$
\langle Y,W'_-(t)\Psi\rangle_-=\langle W_-(t)Y,\Psi\rangle_-,
\quad Y\in{\cal H}_{\alpha,-},\quad
\Psi=(\Psi^0,\Psi^1)\in [S(\Z_-)]^2,
$$
and by $W'_N(t)$ the operator adjoint to $W_N(t)$, $t\in\R$,
$$
\langle Y,W'_N(t)\Psi\rangle_N=\langle W_N(t)Y,\Psi\rangle_N,
\quad Y\in{\cal H}_{\alpha,N},\quad
\Psi=(\Psi^0,\Psi^1)\in [S(\Z_N)]^2,
$$
Here $S(\Z_-)$ denotes the class of rapidly decreasing sequences in $\Z_-$,
$\langle\cdot,\cdot\rangle_-$ stands for the inner product in ${\cal H}_{0,-}$
or for its different extensions.
Similarly, $\langle\cdot,\cdot\rangle_N$ stands for the inner product in ${\cal H}_{0,N}$.
Below we also use the notation $\langle\cdot,\cdot\rangle$  for the inner product in ${\cal H}_{0}$
or for its different extensions.
Write
\beqn\label{2.19}
{\bf g}_-(n,t)=(W'_-(t) Y_0)(n)\quad \mbox{with }\,\,\,Y_0(n)=(\delta_{(-1)n},0),\quad n\le0,\\
{\bf g}_N(n,t)=(W'_N(t) Y_0)(n)\quad \mbox{with }\,\,\,Y_0(n)=(\delta_{(N+1)n},0),\quad n\ge N.
\eeqn
Let ${\bf G}^j_{-,n}(k)$, ${\bf G}^j_{N,n}(k)$, $j=0,1$, denote the vector valued functions
\beqn\label{Phi-j}
{\bf G}^j_{-,n}(k)=
\int_0^{+\infty} {\cal N}_{n0}^{(j)}(s)\, {\bf g}_-(k,-s)\,ds,\quad k\le 0,\quad n=0,\dots,N,\\
{\bf G}^j_{N,n}(k)=
\int_0^{+\infty} {\cal N}_{nN}^{(j)}(s)\, {\bf g}_N(k,-s)\,ds,\quad k\ge N,\quad n=0,\dots,N,
\eeqn
where  ${\cal N}_{nk}(s)$
are defined in (\ref{N}).
 Note that  ${\bf g}_-(0,t)={\bf g}_N(N,t)=0$ and ${\bf G}^j_{-,n}(0)={\bf G}^j_{N,n}(N)=0$
 for any $n\in\{0,\dots,N\}$.
 Introduce
 \be\label{2.17}
\overline{\bf G}^j_n(k)=
\left\{\ba{ll}
\gamma_-{\bf G}^j_{-,n}(k)&\mbox{for }\,\, k\le-1, \\
0&\mbox{for }\,\, k=0,\dots N\\
\gamma_+{\bf G}^j_{N,n}(k)&\mbox{for }\,\, k\ge N+1,
\ea\right|\quad n=0,\dots,N,\quad j=0,1.
\ee
By (\ref{2.21}) and (\ref{2.17}), we have
\be\label{3.10}
\langle \overline{W}(t) Y_0(\cdot),\overline{\bf G}^j_n(\cdot)\rangle
=\gamma_-\langle W_-(t) Y_0,{\bf G}^j_{-,n}\rangle_-
+\gamma_+\langle W_N(t) Y_0,{\bf G}^j_{N,n}\rangle_N,\quad t\ge0.
\ee

Set
$r^{(0)}(n,t)=r(n,t)$, $r^{(1)}(n,t)=\dot r(n,t)$.
\begin{lemma} \label{pro-q}
Let $Y_0\in{\cal H}_{\alpha}$, $\alpha>3/2$, condition {\bf C} or ${\bf C}_0$ hold,
 and  $r(n,t)$ be a solution of problem (\ref{2.34}).
 Then for $t>0$, $j=0,1$, $n=0,1,\dots,N$,
\beqn\label{5.4}
r^{(j)}(n,t)= \langle \overline{W}(t) Y_0(\cdot),\overline{\bf G}^j_n(\cdot)\rangle+\delta_n^j(t),
\quad
 \sup_{n=0,1,\dots,N}|\delta_n^j(t)|\le C\langle t\rangle^{-\beta/2}\Vert Y_0\Vert_{\alpha},
\eeqn
where the operator
$\overline{W}(t)$ is defined in  (\ref{2.21}),
$\beta$ is introduced in Theorem~\ref{l3.1}.
\end{lemma}
{\bf Proof}\, Using (\ref{2.34}) and the bound~(\ref{NN}), we obtain for $t>0$
\be\label{9}
r^{(j)}(n,t)=
\int_0^t \left(
{\cal N}^{(j)}_{n0}(\tau) \gamma_-z_- (-1,t-\tau)+
{\cal N}^{(j)}_{nN}(\tau) \gamma_+z_N(N+1,t-\tau)\right)\,d\tau
+ O(\langle t\rangle^{-\beta/2}).
\ee
We estimate the fist term in the r.h.s. of (\ref{9}).
The second term is estimated by a similar way.
The bounds~(\ref{ubound}) and (\ref{NN}) give
$$
\Big|\int_t^{+\infty} {\cal N}^{(j)}_{n0}(\tau) z_-(-1,t-\tau)\,d\tau\Big|
 \le C\int_t^{+\infty} \langle\tau\rangle^{-\beta/2} \langle t-\tau\rangle^{-3/2}
 \Vert Y_0\Vert_{\alpha}\,d\tau
\le C\langle t\rangle^{-\beta/2}\Vert Y_0\Vert_{\alpha}.
$$
Using (\ref{2.19}) and the equality $W_-(t-\tau)=W_-(t)W_-(-\tau)$, we have
$$
 z_-(-1,t-\tau)=\Big\langle \left(W_-(t-\tau)Y_0\right)(k),
 \left(\ba{l}\delta_{-1 k}\\0\ea\right)\Big\rangle_-
 =\langle W_-(t)Y_0(\cdot), {\bf g}_{-}(\cdot,-\tau)\rangle_-.
$$
Hence,
\begin{eqnarray*}
r^{(j)}(n,t) &=& \int_0^{+\infty}
 \left(
{\cal N}^{(j)}_{n0}(\tau) \gamma_-z_- (-1,t-\tau)+
{\cal N}^{(j)}_{nN}(\tau) \gamma_+z_N(N+1,t-\tau)\right)\,d\tau
+ O(\langle t\rangle^{-\beta/2})\\
&=&\gamma_-\langle \left(W_-(t)Y_0\right)(\cdot), {\bf G}^j_{-,n}(\cdot)\rangle_-
+\gamma_+\langle \left(W_N(t)Y_0\right)(\cdot), {\bf G}^j_{N,n}(\cdot)\rangle_N
+ O(\langle t\rangle^{-\beta/2}).
\end{eqnarray*}
Together with (\ref{3.10}), this implies the representation~(\ref{5.4}). \bo
\medskip

Now we estimate the first term in the r.h.s. of (\ref{5.4}).
\begin{lemma}
Let $Y_0\in{\cal H}_{\alpha}$, $\alpha>3/2$, condition {\bf C} or ${\bf C}_0$ hold.
  Then for $n=0,1,\dots,N$,
\beqn\label{50.4}
| \langle \overline{W}(t) Y_0(\cdot),\overline{\bf G}^j_n(\cdot)\rangle|\le
C\langle t\rangle^{-\beta/2}\Vert Y_0\Vert_{\alpha},\quad t>0,\quad j=0,1.
\eeqn
\end{lemma}
{\bf Proof}\, At first, we prove the following estimate
\begin{equation}\label{50.5}
  \Vert\left(W'_-(t){\bf G}^j_{-,n}\right)(\cdot)\Vert_{-\alpha,-}+
   \Vert\left(W'_N(t){\bf G}^j_{N,n}\right)(\cdot)\Vert_{-\alpha,N}
   \le
C\langle t\rangle^{-\beta/2}
\end{equation}
for any $n=0,1,\dots,N$.
Indeed, applying (\ref{2.19}), we have
$W'_-(t){\bf g}_- (k,-s)={\bf g}_- (k,t-s)$. Hence,
by (\ref{Phi-j}),  (\ref{ubound}) and (\ref{NN}), we obtain
\begin{eqnarray*}
 \Vert\left(W'_-(t){\bf G}^j_{-,n}\right)(\cdot)\Vert_{-\alpha,-}
  &\le&\int_0^{+\infty}|{\cal N}_{n0}^{(j)}(s)|\Vert{\bf g}_-(\cdot,t-s)\Vert_{-\alpha,-}\,ds\\
  &\le&
  C\int_0^{+\infty}\langle s\rangle^{-\beta/2}
  \langle t-s\rangle^{-3/2}\,ds
  \le C_1\langle t\rangle^{-\beta/2}.
\end{eqnarray*}
The similar estimate is true for
$ \Vert\left(W'_N(t){\bf G}^j_{N,n}\right)(\cdot)\Vert_{-\alpha,N}$.
Since
\beqn\nonumber
\left|\langle W_-(t) Y_0,\gamma_-{\bf G}^j_{-,n}\rangle_-\right|
\le\gamma_-\Vert Y_0\Vert_{\alpha,-}\Vert W'_-(t){\bf G}^j_{-,n}\Vert_{-\alpha,-},\\
\left|\langle W_N(t) Y_0,\gamma_+{\bf G}^j_{N,n}\rangle_N\right|
\le\gamma_+\Vert Y_0\Vert_{\alpha,N}\Vert W'_N(t){\bf G}^j_{N,n}\Vert_{-\alpha,N}, \nonumber
\eeqn
the bound~(\ref{50.4}) follows from (\ref{3.10}) and (\ref{50.5}).
\bo
\medskip

Introduce a vector-valued function $\overline{\bf \Gamma}^{j}(n,k)$,  $j=0,1$, by the rule
\beqn\label{Pi-0}
\overline{\bf \Gamma}^j(n,k)=\left\{\ba{lll}
\gamma_-\int_0^{+\infty}\Gamma^-_n(s)\Big(W'_-(-s){\bf G}^j_{-,0}\Big)(k)\,ds,&\mbox{if }\,n\le-1,&k \le-1,\\
\gamma_+\int_0^{+\infty}\Gamma^-_n(s)\Big(W'_N(-s){\bf G}^j_{N,0}\Big)(k)\,ds,&\mbox{if }\,n\le-1,&k \ge N+1,\\
\gamma_-\int_0^{+\infty}\Gamma^+_{n-N}(s)\Big(W'_-(-s){\bf G}^j_{-,N}\Big)(k)\,ds,&\mbox{if }\,n\ge N+1,& k\le-1,\\
\gamma_+\int_0^{+\infty}\Gamma^+_{n-N}(s)\Big(W'_N(-s){\bf G}^j_{N,N}\Big)(k)\,ds,&\mbox{if }\,n\ge N+1,& k\ge N+1,\\
\overline{\bf G}^j_n(k),&\mbox{if }\,n=0,\dots,N,& k\in\Z,\\
0 & \mbox{otherwise}.
\ea\right.
\eeqn
Now we study the large time behavior of $r(n,t)$ for $n\ne0,1\dots,N$.
\begin{lemma} \label{pro-qx}
Assume that $Y_0\in{\cal H}_{\alpha}$, $\alpha>3/2$,
and condition {\bf C} or ${\bf C}_0$ hold. Then the
 solution $r(n,t)$ with $n\not\in\{0,\dots,N\}$ of  problem~(\ref{b.1}), (\ref{b.1+}), (\ref{b.3})
admits the following representation
\be\label{qj}
r^{(j)}(n,t)
=\langle \left(\overline{W}(t) Y_0\right)(\cdot),\overline{\bf \Gamma}^j(n,\cdot)\rangle  +\delta_j(n,t),
\quad j=0,1,\quad t>0,
\ee
where
$\Vert \delta_j(\cdot,t)\Vert_{-\alpha}\le C\langle t\rangle^{-\beta/2}\Vert Y_0\Vert_{\alpha}$.
\end{lemma}
{\bf Proof}\,
We consider the case $n\le -1$ only. For $n\ge N+1$, the proof is similar.
By (\ref{qxt}) and (\ref{5.4}),
\be\label{6.1}
r^{(j)}(n,t)=
\int_{0}^t \Gamma^-_n(t-s)\langle \overline{W}(s)Y_0,\overline{\bf G}^{j}_0\rangle\,ds
+\delta'_j(n,t)\quad\mbox{for }\,\,n\le-1,
\ee
where $\Vert \delta'_j(\cdot,t)\Vert_{-\alpha,-}
\le C\langle t\rangle^{-\beta/2}\Vert Y_0\Vert_{\alpha}$.
Indeed, by (\ref{5.4}) and (\ref{boundK}),
\beqn
\Vert \delta'_j(\cdot,t)\Vert_{-\alpha,-}\!&=&\!
\Big\Vert \int_{0}^t \Gamma^-_n(t-s) \delta_0^j(s)\,ds\Big\Vert_{-\alpha,-}
\le\int_0^t \left\Vert \Gamma^-_n(t-s)\right\Vert_{-\alpha,-}|\delta_0^j(s)|ds
\nonumber\\
\!&\le&\! C\int_0^t (1+t-s)^{-3/2}(1+s)^{-\beta/2}\,ds\Vert Y_0\Vert_{\alpha}
\le C_1\langle t\rangle^{-\beta/2}\Vert Y_0\Vert_{\alpha}.
\nonumber
\eeqn
The first term in the r.h.s. of (\ref{6.1}) has a form
\be\label{4.14}
\int_{0}^t \Gamma^-_n(s)\langle \overline{W}(t-s)Y_0,\overline{\bf G}^{j}_0\rangle\,ds=
\int_{0}^{+\infty} \Gamma^-_n(s)\langle \overline{W}(t-s)Y_0,\overline{\bf G}^{j}_0\rangle\,ds
+\delta''_j(n,t),\quad n\le-1,
\ee
where, by definition,
$ \delta''_j(n,t)=-\ds\int_t^{+\infty}\!\!\!\Gamma^-_n(s)
\langle \overline{W}(t-s)Y_0,\overline{\bf G}^j_0\rangle\, ds$.
The bounds~(\ref{boundK}) and (\ref{50.4}) yield
\be\label{6.5}
\left\Vert \delta''_j(\cdot,t)\right\Vert_{-\alpha,-}\le
\int_t^{+\infty}\!\left\Vert \Gamma^-_n(s)\right\Vert_{-\alpha,-}
\left|\langle \overline{W}(t-s)Y_0,\overline{\bf G}^j_0\rangle\right|ds
\le C\langle t\rangle^{-\beta/2}\Vert Y_0\Vert_{\alpha}.
\ee
Finally, applying (\ref{3.10}) we obtain
\beqn\label{2.23}
&&\int_{0}^{+\infty} \Gamma^-_n(s)\langle \overline{W}(t-s)Y_0,\overline{\bf G}^{j}_0\rangle ds
=
\int_{0}^{+\infty} \Gamma^-_n(s)\langle W_-(t\!-\!s)Y_0,\gamma_-{\bf G}^{j}_{-,0}\rangle_-\,ds\nonumber\\
&&\qquad\qquad \qquad\qquad\qquad\qquad\qquad\quad
+
\int_{0}^{+\infty} \Gamma^-_n(s)\langle W_N(t\!-\!s)Y_0,\gamma_+{\bf G}^{j}_{N,0}\rangle_N\,ds\nonumber\\
&&=\int_{0}^{+\infty} \Gamma^-_n(s)\langle W_-(t)Y_0,\gamma_-W'_-(-s){\bf G}^{j}_{-,0}\rangle_-\,ds+
\int_{0}^{+\infty}\Gamma^-_n(s)
\langle W_N(t)Y_0,\gamma_+W'_N(-s){\bf G}^{j}_{N,0}\rangle_N\,ds
\nonumber\\
&&=\langle W_-(t)Y_0,\overline{\bf \Gamma}^j(n,\cdot)\rangle_-+
\langle W_N(t)Y_0,\overline{\bf \Gamma}^j(n,\cdot)\rangle_N=
\langle \overline{W}(t)Y_0,\overline{\bf \Gamma}^j(n,\cdot)\rangle.
\nonumber
\eeqn
Hence, the  bounds~(\ref{6.1})--(\ref{6.5}) imply (\ref{qj})
with $\delta_j(n,t)=\delta'_j(n,t)+\delta''_j(n,t)$ for $n\le-1$.
\bo
\begin{remark}
We put $\overline{\bf \Gamma}^j(n,k)=\overline{\bf G}^j_n(k)$
for $n=0,1,\dots,N$. Hence, the representation~(\ref{5.4})
is a particular case of (\ref{qj}).
\end{remark}

Now we estimate the first term in the r.h.s. of (\ref{qj}).
\begin{lemma}\label{l3.5}
Assume that $Y_0\in{\cal H}_{\alpha}$, $\alpha>3/2$,
and condition {\bf C} or ${\bf C}_0$ hold. Then
\be\label{3.14}
\Vert
\langle \left(\overline{W}(t) Y_0\right)(k),\overline{\bf \Gamma}^j(\cdot,k)\rangle\Vert_{-\alpha}\le
C\langle t\rangle^{-\beta/2}\Vert Y_0\Vert_{\alpha},\quad t\ge0,\quad j=0,1.
\ee
\end{lemma}
{\bf Proof}\, For any $n$, we have
\begin{equation}\label{3.18}
\left|\langle \left(\overline{W}(t) Y_0\right)(\cdot),\overline{\bf \Gamma}^j(n,\cdot)\rangle\right|
\le \Vert Y_0\Vert_{\alpha}
\left(\Vert W'_-(t)\overline{\bf \Gamma}^j(n,\cdot)\Vert_{-\alpha,-}
+\Vert W'_N(t)\overline{\bf \Gamma}^j(n,\cdot)\Vert_{-\alpha,N}\right).
\end{equation}
Using (\ref{Pi-0}), we estimate the first term in the r.h.s. of (\ref{3.18}):
\begin{eqnarray*}
\Vert W'_-(t)\overline{\bf \Gamma}^j(n,\cdot)\Vert_{-\alpha,-} \le
\left\{\begin{array}{ll}
\ds\gamma_-\int_0^{+\infty}|\Gamma^-_n(s)|\,\Vert W'_-(t-s){\bf G}^j_{-,0}\Vert_{-\alpha,-}\,ds&\mbox{for }\,n\le-1,\\
\ds\gamma_-\int_0^{+\infty}|\Gamma^+_{n-N}(s)|\,\Vert W'_-(t-s){\bf G}^j_{-,N}\Vert_{-\alpha,-}\,ds&\mbox{for }\,n\ge N+1,\\
\gamma_-\Vert W'_-(t){\bf G}^j_{-,n}\Vert_{-\alpha,-}&\mbox{for }\,n=0,1,\dots,N.
\end{array}
\right.\end{eqnarray*}
Hence, applying (\ref{boundK}) and (\ref{50.5}), we obtain
\be\label{3.16}
\sqrt{\sum_{n\in\Z}\langle n\rangle ^{-2\alpha}
\Vert W'_-(t)\overline{\bf \Gamma}^j(n,\cdot)\Vert^2_{-\alpha,-}}
\le C_N\gamma_-\langle t\rangle^{-\beta/2}.
\ee
The similar bound is valid for
$\Vert\Vert W'_N(t)\overline{\bf \Gamma}^j(n,\cdot)\Vert_{-\alpha,N}\Vert_{-\alpha}$.
The bound (\ref{3.16}) implies (\ref{3.14}). \bo
\medskip

Introduce an operator
 $\Omega:{\cal H}_{\alpha}\to{\cal H}_{-\alpha}$, $\alpha>3/2$, by the rule
\be\label{Omega}
\Omega:Y\to
Y(n)+\Big(\langle Y(\cdot),\overline{\bf \Gamma}^{0}(n,\cdot)\rangle,
m_n\langle Y(\cdot),\overline{\bf \Gamma}^{1}(n,\cdot)\rangle\Big),
\quad n\in\Z.
\ee
It follows from (\ref{3.14}) that the operator $\Omega$ is
 bounded, $\Vert\Omega Y\Vert_{-\alpha}\le C\Vert Y\Vert_{\alpha}$
for any $\alpha>3/2$.
\medskip\\
{\bf Proof of Theorem~\ref{thC}}\,
Theorem~\ref{t1} and Lemmas~\ref{pro-q}--\ref{l3.5}
imply assertions of Theorem~\ref{thC}.
Indeed, the representation~(\ref{2.38}) follows from
(\ref{2.1}), (\ref{2.21}), (\ref{5.4}) and (\ref{qj}).
The bounds~(\ref{ubound}), (\ref{uboundN}), (\ref{50.4})
and (\ref{3.14}) imply the bound~(\ref{2.31'}).\bo

\setcounter{equation}{0}
\section{Appendix A: Case $N=0$}\label{appendixA}
If $N=0$, then there is the unique ``defect'' in the chain,
which is the particle located at origin with mass $m_0$
different, generally speaking, from masses of the other particles and
with the  constant of the external force
$\mu_0$  unequal to $\mu_\pm$, in general.

\subsection{Conditions ${\bf C}$ and ${\bf C}_0$ in the particular cases}

Now  we simplify  conditions ${\bf C}$ and ${\bf C}_0$
for $N=0$ and for some particular cases of the chain.
In Remark~\ref{r2.8}, we consider a particular case~({\bf P1}).
Now we study another two cases.
\smallskip

{\bf Particular case (P2)}:
Assume that the oscillators in the chain have identical masses equal to unity,
 i.e., $m_\pm=m_0=1$, and let, for simplicity, $\kappa_-\le \kappa_+$.
This case was considered in \cite{rjmp-2018}. In this case, conditions ${\bf C}$ and ${\bf C}_0$
are of the following form
 \begin{itemize}
  \item [${\bf C}$]
$\mu_0< K_+(a_-)$ if $a_-\ge a_+$; $\mu_0<K_-(a_+)$ if $a_+\ge a_-$;\\
$\mu_0> K_0(\kappa_-)$ if $\kappa_-\not=0$;\\
$\mu_0> K_-(\kappa_+)$ or  $\mu_0< K_0(a_-)$ if $a_-\le\kappa_+$;\\
$\mu_0\not=0$ if  $\mu_-=\mu_+=0$.
\end{itemize}
For example, $\mu_0\in\left(0,2\max(\nu_-,\nu_+)\sqrt{|\nu_-^2-\nu_+^2|}\right)$
if  $\kappa_\pm=0$ and $\nu_-\not=\nu_+$.
Note that first two restrictions
of condition~{\bf C} exclude the case when  $m_-=m_+$, $\gamma_-=\gamma_+$ and $\mu_-=\mu_+$,
since in this case $K_+(a_-)=K_-(a_+)=\kappa_-^2$ and $K_0(\kappa_-)=\kappa_-^2$ if $\kappa_-\not=0$.
However, the case $\kappa_-=\kappa_+\not=0$ and $\nu_-=\nu_+$ is included in condition~${\bf C}_0$.
\smallskip
  \begin{itemize}
   \item [${\bf C}_0$] One of the following restrictions is fulfilled.

   (i) $a_->a_+$,  $\mu_0=K_+(a_-)$.

(ii)   $a_-<a_+$, $\mu_0=K_-(a_+)$.

(iii)  $a_+=a_-$, $(\kappa_-,\kappa_+)\not=(0,0)$, $\mu_0=\bar\kappa^2$.

(iv)    $\kappa_-\not=0$,  $\mu_0=K_0(\kappa_-)$ (if $K_0(\kappa_-)\ge0$).

(v)  $a_-\le\kappa_+$,  $\mu_0=K_-(\kappa_+)$.

(vi)  $a_-<\kappa_+$,  $\mu_0=K_0(a_-)$ (if $K_0(a_-)\ge0$).
\end{itemize}
\begin{remark}
Condition ${\bf C}_0$ (i)--(iv) includes the following particular cases:
\begin{itemize}
\item{}  $\kappa_-=\kappa_+$, $\nu_-\not=\nu_+$,
   $\mu_0=\kappa_-^2+2\max(\nu_-,\nu_+)\sqrt{|\nu_-^2-\nu_+^2|}$
(see cases (i) and (ii));

\item{}    $\kappa_-=\kappa_+\not=0$, $\mu_0=K_0(\kappa_-)=\bar\kappa^2$
(see case (iv));

\item{}   $\kappa_-=\kappa_+\not=0$, $\nu_-=\nu_+$, $\mu_0=\mu_+$
 (see  cases (iii) and (iv)).
\end{itemize}

\end{remark}

{\bf Particular case (P3)}:
Assume that external forces don't act on the oscillators in the chain
except the defect, i.e.,
$\mu_\pm=0$, $\mu_0>0$. In this case, $\kappa_\pm=0$, $a_\pm=2\nu_\pm$
 and conditions ${\bf C}$ and ${\bf C}_0$  are of a form

\begin{itemize}
  \item [${\bf C}$]
$\mu_0>0$;\\
$\mu_0<4\nu_-^2\left(m_0-\frac{m_-+m_+}{2}\right)
+2m_+\nu_-\sqrt{\nu_-^2-\nu_+^2}$ if $\nu_-\ge \nu_+$;\\
$\mu_0<4\nu_+^2\left(m_0-\frac{m_-+m_+}{2}\right)
+2m_-\nu_+\sqrt{\nu_+^2-\nu_-^2}$ if  $\nu_+\ge \nu_-$.

  \item [${\bf C}_0$] $\mu_0>0$ and one of the following restrictions is fulfilled.

(i)
   $\mu_0=4\nu_-^2\left(m_0-\frac{m_-+m_+}{2}\right)
+2m_+\nu_-\sqrt{\nu_-^2-\nu_+^2}$ if $\nu_-\ge \nu_+$.

(ii)
 $\mu_0=4\nu_+^2\left(m_0-\frac{m_-+m_+}{2}\right)
+2m_-\nu_+\sqrt{\nu_+^2-\nu_-^2}$ if $\nu_-\le \nu_+$.
\end{itemize}
Note that if $\nu_-= \nu_+$, then conditions ${\bf C}$ and ${\bf C}_0$ imply that the mass of the defect satisfies the following restriction $m_0>(m_-+m_+)/2$.

%
%
%
%
%
%
%

\subsection{$\tilde D(\omega\pm i0)$ for $\omega\in\Lambda\setminus\Lambda^0$}
If $N=0$, then
$\tilde D(\omega)$ is of the form (\ref{D-N0}).
We have $\tilde D(\omega)\not=0$ for $\omega\in\CC_\pm$ by Lemma~\ref{l2.A} from Appendix~B.
Now we study $\tilde D(\omega\pm i0)=\lim\limits_{\ve\to+0}\tilde D(\omega\pm i\ve)$
for $\omega\in\Lambda\setminus\Lambda^0$.
\begin{lemma}\label{l8.2}
  $\tilde D(\omega\pm i0)\not=0$ for $\omega\in\Lambda\setminus \Lambda^0$.
\end{lemma}
{\bf Proof}\,
Let $\omega\in\Lambda_-\setminus \Lambda_-^0$ and $\omega\notin\Lambda_+$.
Then $\Re\theta_-(\omega+i0)\in(-\pi,0)\cup(0,\pi)$
and $\Im\theta_-(\omega+i0)=0$.
Moreover, ${\rm sign}(\sin\theta_-(\omega+i0))={\rm sign}\,\omega$.
Therefore,
$$
\ba{lll}
\Im\tilde D(\omega+i0)&=& -\gamma_-\sin\theta_-(\omega+i0)\\
&=&\ds
-\frac{m_-}2\left\{\ba{lll}
{\rm sign}(\omega)\sqrt{\omega^2-\kappa_-^2}\sqrt{a_-^2-\omega^2}\not=0,&\kappa_-\not=0,\\
\omega\sqrt{a_-^2-\omega^2}\not=0,&\kappa_-=0.
\ea\right.\ea
$$
Hence, $\tilde D(\omega+i0)\not=0$ for such values of $\omega$.
Similarly, we can check that for $\omega\in\Lambda_+\setminus \Lambda_+^0$
and $\omega\notin\Lambda_-$, $\Im\tilde D(\omega+i0)\not=0$.
For $\omega\in(\Lambda_-\cap\Lambda_+)\setminus\cup_\pm \Lambda^0_\pm$,
we have
\beqn
\Im\tilde D(\omega+i0)=
-\gamma_-\sin\theta_-(\omega+i0)-\gamma_+\sin\theta_+(\omega+i0)
\quad\quad\quad\quad\quad\quad\quad\quad\quad\quad\nonumber\\
=\!-\frac{1}2\left\{
\ba{lr}
{\rm sign}(\omega)
\sum\limits_\pm m_\pm
\sqrt{\omega^2-\kappa^2_\pm}\sqrt{a_\pm^2-\omega^2},&\!\!\mbox{if } \kappa_\pm\not=0\\
{\rm sign}(\omega)\Big(m_-|\omega|\sqrt{4\nu_-^2\!-\!\omega^2}\!+\!
m_+\sqrt{\omega^2\!-\!\kappa^2_+}\sqrt{a_+^2\!-\!\omega^2}\Big),&\!\!\!\mbox{if } \kappa_-\!=\!0,\kappa_+\not\!=0\\
\omega\sum\limits_\pm m_\pm\sqrt{4\nu_\pm^2-\omega^2},& \!\!\mbox{if } \kappa_-=\kappa_+=0
\ea\right.\nonumber
\eeqn
Therefore, $\Im\tilde D(\omega+i0)\not=0$ for such values of $\omega$.
Since
$\tilde D(\omega-i0)=\overline{\tilde D(\omega+i0)}$
for $\omega\in\Lambda\setminus\Lambda^0$,
$\tilde D(\omega-i0)\not=0$ for $\omega\in\Lambda\setminus\Lambda^0$.
\bo

\subsection{$\tilde D(\omega)$ for $\omega\in\R\setminus \Lambda$}
To prove Theorem~\ref{tB}, we first prove the following lemma,
using the technique of \cite{rjmp-2018}.
 \begin{lemma}\label{l4}
  $\tilde D(\omega)\not=0$ for $\omega\in\R\setminus \Lambda$
  iff the following conditions hold.
  \begin{itemize}
  \item [(i)]
   $\tilde D(\max(a_-,a_+))\le0$.
 \item [(ii)]
 If  $\min(\kappa_-,\kappa_+)>0$, then $\tilde D(\min(\kappa_-,\kappa_+))\ge0$.
 \item [(iii)]
 If  $a_\mp<\kappa_\pm$, then either
 $\tilde D(\kappa_\pm)\ge0$ or $\tilde D(a_\mp)\le0$.
\end{itemize}
\end{lemma}
{\bf Proof}\, To prove this lemma
we apply the following formulas from \cite{rjmp-2018}.

Let $\kappa_+\not=0$, $\omega\in \R$ and $|\omega|<\kappa_+$. Then,
$\Re\theta_+(\omega)=0$ and $e^{i\theta_+(\omega)}=e^{-\Im \theta_+(\omega)}<1$.
Moreover,
for $\omega\in\R:\,|\omega|<\kappa_+$,
\be\label{a}
\nu_+^2(1-e^{i\theta_+(\omega)})=
\frac12\left(\omega^2-\kappa_+^2+\sqrt{\kappa_+^2-\omega^2}\sqrt{a_+^2-\omega^2}\right)=
\frac12(\omega^2+\kappa_-^2)-K_0(\omega),
\ee
where $K_0(\omega)$ is defined in (\ref{K0}).

 If  $\omega\in \R$ and $|\omega|>a_\pm$, then
$e^{i\theta_\pm(\omega)}=-e^{-\Im \theta_\pm(\omega)}$
and
\be\label{b}
\nu_\pm^2(1-e^{i\theta_\pm(\omega)})=
\frac12\left(\omega^2-\kappa_\pm^2-\sqrt{\omega^2-\kappa_\pm^2}\sqrt{\omega^2-a_\pm^2}\right)
=\frac12(\omega^2+\kappa_\mp^2)-K_\pm(\omega),
\ee
where $K_\pm(\omega)$ is defined in (\ref{Kpm}).
Using formulas~(\ref{a}) and (\ref{b}), we rewrite conditions (i)--(iii) from Lemma~\ref{l4}.
\medskip

{\bf Condition (i)}. Let $\max(a_-,a_+)=a_-$.
By (\ref{b}), we have
\beqn
\tilde D(a_-)&=&\mu_0-m_0a_-^2+\gamma_-(1-e^{i\theta_-(a_-)})+\gamma_+(1-e^{i\theta_+(a_-)})
\nonumber\\
&=&\mu_0-m_0a_-^2+2\gamma_-+m_+\left(\frac12(a_-^2+\kappa_-^2)-K_+(a_-)\right).\nonumber
\eeqn
Note that   $\Re\tilde D(\omega_1)<\Re\tilde D(\omega_2)$ for
$|\omega_1|>|\omega_2|\ge \max(a_-,a_+)$,
and  $\Re\tilde D(\omega)\to-\infty$ as $|\omega|\to\infty$.
Therefore, $\tilde D(\omega)\not=0$ for $|\omega|>a_-$ iff $\tilde D(a_-)\le0$.
In turn,  $\tilde D(a_-)\le0$ if and only if
\be\label{4.1}
\mu_0\le \kappa_-^2(m_0-m_+)+4\nu_-^2(m_0-\frac{m_-+m_+}{2})+m_+K_+(a_-),\quad \mbox{where }\,a_-\ge a_+.
\ee
Similarly, if $\max(a_-,a_+)=a_+$, then $\tilde D(a_+)\le0$ iff
\be\label{4.2}
\mu_0\le
\kappa_+^2(m_0-m_-)+ 4\nu_+^2(m_0-\frac{m_-+m_+}{2})+m_-K_-(a_+),\quad \mbox{where }\,a_+\ge a_-.
\ee
If $a_-=a_+$, then (\ref{4.1}) and (\ref{4.2}) become
$$
\mu_0\le m_0(\mu_-+4\gamma_-)/m_-+2\gamma_-+2\gamma_+.
$$

{\bf Condition (ii)}. Let $\min(\kappa_-,\kappa_+)=\kappa_-$.
By (\ref{a}), we have
$$
\tilde D(\kappa_-)=\mu_0-m_0\kappa_-^2+\gamma_+(1-e^{i\theta_+(\kappa_-)})=
\mu_0-m_0\kappa_-^2+m_+(\kappa_-^2-K_0(\kappa_-)).
$$
 Note that   $\Re\tilde D(\omega_1)>\Re\tilde D(\omega_2)$ for
$|\omega_1|<|\omega_2|< \min(\kappa_-,\kappa_+)$,
and  $\Re\tilde D(0)>0$.
Then, $\tilde D(\omega)\not=0$ for $|\omega|<\kappa_-$ iff $\tilde D(\kappa_-)\ge0$.
In turn,
 $\tilde D(\kappa_-)\ge0$ iff
\be\label{4.3}
\mu_0\ge\kappa_-^2(m_0-m_+)+m_+ K_0(\kappa_-),\quad \mbox{where }\,\kappa_-\not=0.
\ee

{\bf Condition (iii)}. Let  $a_-<\kappa_+$.
Note that   $\Re\tilde D(\omega_1)<\Re\tilde D(\omega_2)$ for
$|\omega_1|>|\omega_2|$.
Therefore,
 $\tilde D(\omega)\not=0$ for $a_-<|\omega|<\kappa_+$ iff either
 $\tilde D(\kappa_+)\ge0$  or $\tilde D(a_-)\le0$.
 then
$\tilde D(\kappa_+)\ge0$ or $\tilde D(a_-)\le0$.
Using (\ref{b}), we have
\beqn
\tilde D(\kappa_+)=\mu_0-m_0\kappa_+^2+\gamma_-(1-e^{i\theta_-(\kappa_+)})
=\mu_0-m_0\kappa_+^2+m_-\left(\kappa_+^2-K_-(\kappa_+)\right).\nonumber
\eeqn
Hence,
\be\label{4.4}
\tilde D(\kappa_+)\ge0\,\, \Longleftrightarrow\,\,
\mu_0\ge\kappa_+^2(m_0-m_-)+m_- K_-(\kappa_+),\quad \mbox{where }\,a_-<\kappa_+.
\ee
Secondly, using (\ref{a}), we have
\beqn
\ba{lll}
\tilde D(a_-)&=&\mu_0-m_0a_-^2+2\gamma_-+\gamma_+(1-e^{i\theta_+(a_-)})\\
&=&\mu_0-m_0(\kappa_-^2+4\nu_-^2)+2m_-\nu_-^2+m_+\left(\kappa_-^2+2\nu_-^2-K_0(a_-)\right).\nonumber
\ea\eeqn
Therefore,
\beqn\label{4.4'}
\tilde D(a_-)\le0 \Longleftrightarrow
\mu_0\le\kappa_-^2(m_0\!-\!m_+)\!+\! 4\nu_-^2\Big(m_0-\frac{m_-\!+\!m_+}{2}\Big)\!+\!m_+K_0(a_-).\bo
\eeqn
\begin{remark}
The formulas (\ref{4.1})--(\ref{4.4'})
 are the part of the restrictions on the constants in conditions ${\bf C}$ and ${\bf C}_0$.
 If restrictions~(\ref{4.1})--(\ref{4.4'}) are not fulfilled, then
 there is a point $\omega_*\in\R\setminus\{\Lambda\cup0\}$ such that
 $\tilde D(\omega_*)=0$. Furthermore, $\tilde D'(\omega_*)\not=0$, since
 $$
 \tilde D'(\omega_*)=-2|\omega_*|\left(m_0+\sum\limits_\pm\gamma_\pm
 e^{-\Im\theta_\pm(\omega_*)} |\kappa^2_\pm-\omega_*^2|^{-1/2}|a^2_\pm-\omega_*^2|^{-1/2}\right).
 $$
 Hence, $\tilde {\cal N}(\omega)$ has simple poles
 at points $\omega=\pm\omega_*$. Then, the bound~(\ref{NN}) does not hold.
\end{remark}

\subsection{Asymptotics of $\tilde D(\omega)$ near points in $\Lambda^0$}

Now we study the asymptotic behavior of $\tilde D(\omega)$
near singular points in $\Lambda^0$ using the following formulas
\be\label{a8}
e^{i\theta_\pm(\omega)}= 1+\frac{i}{\nu_\pm}\sqrt{\omega^2-\kappa_\pm^2}-
\frac1{2\nu_\pm^2}(\omega^2-\kappa_\pm^2)
+\dots,
\,\,\,\omega\to \kappa_\pm,
\ee
where $\omega\in\mathbb{C}_+$,
 $\Im(\sqrt{\omega^2-\kappa_\pm^2})>0$. Here
${\rm sign}(\Re\sqrt{\omega^2-\kappa_\pm^2})={\rm sign}(\Re\omega)$
for $\omega\in\mathbb{C}_+$.
 This choice of the branch of the complex root  $\sqrt{\omega^2-\kappa_\pm^2}$ follows
 from the condition  $\Im\theta_\pm(\omega)>0$.
Similarly,
\be\label{a9}
e^{i\theta_\pm(\omega)}=-1+\frac{i}{\nu_\pm}\sqrt{a_\pm^2-\omega^2}+\frac1{2\nu_\pm^2}(a^2-\omega^2)
-\frac{i}{8\nu_\pm^3}(a_\pm^2-\omega^2)^{3/2}+\dots
\ee
for $\omega\to a_\pm$, $\omega\in\mathbb{C}_+$.
Here the branch of the complex root $\sqrt{a_\pm^2-\omega^2}$ is chosen so that
 ${\rm sign}(\Re\sqrt{a_\pm^2-\omega^2})={\rm sign}(\Re\omega)$,
by the condition $\Im\theta_\pm(\omega)>0$.
If $\kappa_\pm=0$, then
\beqn\label{a10}
e^{i\theta_\pm(\omega)}=\left\{
\ba{lll}
1+\frac{i\omega}{\nu_\pm}-\frac{\omega^2}{2\nu_\pm^2}-\frac{i\omega^3}{8\nu_\pm^3}+\dots
& \mbox{for }& \omega\to0\\
-1+i\sqrt{4\nu_\pm^2-\omega^2}/\nu_\pm+\dots& \mbox{for }& \omega\to 2\nu_\pm
\ea\right|\quad \omega\in\mathbb{C}_+.
\eeqn
\begin{lemma}\label{lemmaA}
Let $\omega_0\in\Lambda^0$ and $\omega_0\not=0$. If condition {\bf C} holds, then
 \be\label{lemmaA1}
\tilde {\cal N}(\omega)=C_1+iC_2(\omega^2-\omega_0^2)^{1/2}+\dots\,\,\,\mbox{as }\,\,
\omega\to\pm\omega_0,\,\,\,\,\omega\in\mathbb{C}_\pm,
\ee
 with some $C_1, C_2\not=0$.
If   condition ${\bf C}_0$ holds, then either (\ref{lemmaA1}) is true or
 \be\label{lemmaA2}
\tilde {\cal N}(\omega)=iC_1(\omega^2-\omega_0^2)^{-1/2}+C_2+\dots\,\,\,\mbox{as }\,\,\omega\to\pm\omega_0,
\quad\omega\in\mathbb{C}_\pm,
\ee
 with some $C_1\not=0$.

 Let $\omega_0=0\in\Lambda^0$ and conditions {\bf C} or ${\bf C}_0$ hold.
 Then
 \be\label{lemmaA3}
\tilde {\cal N}(\omega)=C_1+iC_2\omega+\dots\,\,\,\mbox{as }\,\,
\omega\to0,\quad \omega\in\mathbb{C}_\pm,
\ee
 with some $C_1, C_2\not=0$.
\end{lemma}
{\bf Proof}\,
{\bf(1)} Let $\omega_0=\pm\kappa_-$. We consider the following cases:
{\bf(1.1)} $\kappa_-=\kappa_+=0$; {\bf(1.2)} $\kappa_-=\kappa_+\not=0$;
{\bf(1.3)} $\kappa_-=0$, $\kappa_+>0$; {\bf(1.4)} $0<\kappa_-<\kappa_+$.
\medskip

{\bf(1.1)} If $\kappa_-=\kappa_+=0$, then
we apply the representation (\ref{a10}) to $e^{i\theta_\pm(\omega)}$ and obtain
\beqn
\tilde D(\omega)=\mu_0-i\omega(\sqrt{m_-\gamma_-}+\sqrt{m_+\gamma_+})
+\dots\,\,\,\mbox{as }\,\,\omega\to0.\nonumber
\eeqn
In particular,
$\tilde D(0)=\mu_0\not=0$ by condition {\bf C} or ${\bf C}_0$.
Hence, (\ref{lemmaA3}) is true with
$C_1=\mu_0^{-1}$.
Note that if $\mu_\pm=\mu_0=0$
(this case is excluded by conditions {\bf C} and ${\bf C}_0$),
then $\tilde {\cal N}(\omega)$ has a simple pole at zero,
\be\label{5.36'}
\tilde {\cal N}(\omega)=\frac{i}{\omega(\sqrt{m_-\gamma_-}+\sqrt{m_+\gamma_+})}
+\frac{m_-+m_+-2m_0}{2(\sqrt{m_-\gamma_-}+\sqrt{m_+\gamma_+})^2}+\dots,\,\,\,\omega\to0.
\ee


{\bf(1.2)} If $\kappa_-=\kappa_+\not=0$, then
we apply the representation (\ref{a8}) to $e^{i\theta_\pm(\omega)}$ and obtain
$$
\tilde D(\omega)=\tilde D(\pm\kappa_-)
-i(\sqrt{m_-\gamma_-}+\sqrt{m_+\gamma_+})(\omega^2-\kappa_-^2)^{1/2}+\dots
\,\,\mbox{as }\,\, \omega\to\pm\kappa_-,\,\, \omega\in\mathbb{C}_+,
$$
where  $\tilde D(\pm\kappa_-)=\mu_0-m_0\kappa_-^2\not=0$
 iff  $\mu_0\not=m_0\kappa_-^2$ (or $\ds \frac{\mu_0}{m_0}\not=\frac{\mu_-}{m_-}$).
 Hence, if condition {\bf C} holds, then (\ref{lemmaA1})  is true with
  $\omega_0=\pm\kappa_-$, $C_1=(\tilde D(\pm\kappa_-))^{-1}$.
If condition ${\bf C}_0$~{\bf(iv)} holds, then
 (\ref{lemmaA2}) is true.
\medskip

{\bf(1.3)} If $\kappa_-=0$ and $\kappa_+>0$, then
the function $e^{i\theta_+(\omega)}$
is an analytic function in a small neighborhood of origin.
Applying the representation (\ref{a10}) to $e^{i\theta_-(\omega)}$, we obtain
$$
\tilde D(\omega)=\tilde D(0)+c_1\omega+c_2\omega^2+\dots,\,\,\,\omega\to0,
$$
with $\tilde D(0)=\mu_0+\gamma_+\Big(1- e^{-\Im\theta_+(0)}\Big)$ and
 some  constants $c_1$, $c_2$. $\tilde D(0)>0$,
 since
\be\label{3.17}
 e^{-\Im\theta_+(0)}=4\left(\frac{\kappa_+}{\nu_+}
 +\sqrt{4+\frac{\kappa^2_+}{\nu^2_+}}\right)^{-2}<1\quad \mbox{if }\,\,\kappa_+>0.
\ee
Hence, (\ref{lemmaA3}) holds with
 $C_1=(\tilde D(0))^{-1}>0$.
\medskip

{\bf(1.4)} If $0<\kappa_-<\kappa_+$, then
$e^{i\theta_+(\omega)}$ is an analytic function
in a small neighborhood of the points $\omega=\pm\kappa_-$.
Applying the representation (\ref{a8}) to $e^{i\theta_-(\omega)}$, we obtain
$$
\tilde D(\omega)=
\tilde D(\pm\kappa_-)-i\sqrt{m_-\gamma_-}(\omega^2-\kappa_-^2)^{1/2}+O(|\omega\mp\kappa_-|)
\quad \mbox{as }\,\,\omega\to\pm\kappa_-,
$$
$\omega\in\mathbb{C}_+$, and  (\ref{a}) gives
$$
\tilde D(\pm\kappa_-)  =\mu_0-m_0\kappa^2_-+m_+(\kappa_-^2-K_0(\kappa_-))
=\mu_0-\kappa^2_-(m_0-m_+)-m_+K_0(\kappa_-).
$$
Hence,
$$
\tilde D(\pm\kappa_-)\not=0\quad \Longleftrightarrow\quad
\mu_0\not=\kappa^2_-(m_0-m_+)+m_+K_0(\kappa_-).
$$
Therefore,
if condition {\bf C} holds, then (\ref{lemmaA1}) is true with  $\omega_0=\pm\kappa_-$.
If condition ${\bf C}_0$~{\bf(iv)} holds, then (\ref{lemmaA2}) is true.
\medskip

{\bf(2)} Let $\omega_0=\pm\kappa_+$. There are five cases:
{\bf(2.1)} $\kappa_+>a_-$; {\bf(2.2)} $\kappa_+=a_-$;
{\bf(2.3)} $\kappa_-<\kappa_+<a_-$;
{\bf(2.4)} $\kappa_-=\kappa_+=0$ (see case {\bf(1.1)});
{\bf(2.5)} $\kappa_-=\kappa_+\not=0$ (see case {\bf(1.2)}).
\medskip

{\bf(2.1)} Let  $\kappa_+>a_-$. Then
the function $e^{i\theta_-(\omega)}=- e^{-\Im\theta_-(\omega)}$ is analytic
in a small neighborhood of the points $\omega_0=\pm\kappa_+$.
Applying the representation (\ref{a8}) to $e^{i\theta_+(\omega)}$, we obtain
$$
\tilde D(\omega)
=\tilde D(\pm\kappa_+)-i\sqrt{m_+\gamma_+}(\omega^2-\kappa_+^2)^{1/2}+O(|\omega\mp\kappa_+|)
\quad \mbox{as }\,\,\omega\to\pm\kappa_+,\quad\omega\in\mathbb{C}_+.
$$
Using (\ref{b}), we obtain
$$
\tilde D(\pm\kappa_+)=\mu_0-m_0\kappa^2_++\gamma_-(1+e^{-\Im\theta_-(\kappa_+)})
=\mu_0-\kappa^2_+(m_0-m_-)-m_-K_-(\kappa_+).
$$
Hence,
$$
\tilde D(\pm\kappa_+)\not=0\quad \Longleftrightarrow\quad
\mu_0\not=\kappa^2_+(m_0-m_-)+m_-K_-(\kappa_+).
$$
Therefore, condition {\bf C} implies  (\ref{lemmaA1})  with  $\omega_0=\pm\kappa_+$,
 $C_1=(\tilde D(\pm\kappa_+))^{-1}$.
If condition ${\bf C}_0$~{\bf(v)} holds, then
(\ref{lemmaA2}) is true.
\medskip

{\bf(2.2)} If $\kappa_+=a_-$, then
we apply (\ref{a8}) and (\ref{a9}) to $e^{i\theta_+(\omega)}$
and $e^{i\theta_-(\omega)}$, respectively,
 and obtain
$$
\tilde D(\omega)=\tilde D(\pm\kappa_+)-(\kappa_+^2-\omega^2)^{1/2}(\sqrt{m_+\gamma_+}+i\sqrt{m_-\gamma_-})+
O(|\omega\mp\kappa_+|),\,\,\,\omega\to\pm\kappa_+,
$$
$\omega\in\mathbb{C}_+$.
Here $\tilde D(\pm\kappa_+)=\mu_0-m_0\kappa_+^2+2\gamma_-$.
Therefore, condition~{\bf C} implies (\ref{lemmaA1})
with $\omega_0=\pm\kappa_+$.
If condition ${\bf C}_0$~{\bf(v)} holds, then (\ref{lemmaA2}).
\medskip

{\bf(2.3)} If $\kappa_+\in(\kappa_-,a_-)$, then
we apply (\ref{a8}) to $e^{i\theta_+(\omega)}$ and obtain
$$
\tilde D(\omega)=
\tilde D(\kappa_+\pm i0)-i\sqrt{m_+\gamma_+}(\omega^2-\kappa_+^2)^{1/2}+
O(|\omega-\kappa_+|),\,\,\,\omega\to\kappa_+\pm i0,
$$
where
$\tilde D(\kappa_+\pm i0)=\mu_0-m_0\kappa_+^2+\gamma_-
\left(1- e^{\pm i\theta_-(\kappa_++i0)}\right)$.
Therefore,
$$
\ba{lll}
\Im\tilde D(\kappa_+\pm i0)&=&\mp\gamma_-\sin\theta_-(\kappa_++i0)\nonumber\\
&=&\mp\frac{m_-}{2}\left\{
\ba{lll}
\sqrt{\kappa_+^2-\kappa^2_-}\sqrt{a_-^2-\kappa_+^2},&\mbox{if }\, \kappa_->0\\
\kappa_+\sqrt{a_-^2-\kappa_+^2},&\mbox{if }\, \kappa_-=0
\ea
\right.\ea
$$
Then, $\Im\tilde D(\kappa_+\pm i0)\not=0$, and (\ref{lemmaA1}) follows.
\medskip

{\bf(3)} Let $\omega_0=\pm a_-$. There are the following cases:
{\bf(3.1)} $a_->a_+$; {\bf(3.2)} $a_-=a_+$; {\bf(3.3)} $a_-\in(\kappa_+,a_+)$;
{\bf(3.4)} $a_-=\kappa_+$ (see the case {\bf (2.2)}); {\bf(3.5)} $a_-<\kappa_+$.
\medskip

{\bf(3.1)} If  $a_->a_+$, then
the function $e^{i\theta_+(\omega)}= -e^{-\Im\theta_+(\omega)}$ is analytic
in a small neighborhood of the points $\omega_0=\pm a_-$.
We apply (\ref{a9}) to $e^{i\theta_-(\omega)}$ and obtain
$$
\tilde D(\omega)=\tilde D(\pm a_-)-i\sqrt{m_-\gamma_-}(a_-^2-\omega^2)^{1/2}+O(|\omega\mp a_-|),
\,\,\,\omega\to\pm a_-,\,\,\omega\in\mathbb{C}_+,
$$
where (see (\ref{b}))
\beqn
\ba{lll}
\tilde D(a_-)&=&\mu_0-m_0a_-^2+2\gamma_-+\gamma_+(1-e^{i\theta_+(a_-)})\\
&=&\mu_0-m_0(\kappa_-^2+4\nu_-^2)+2m_-\nu_-^2+m_+\left(\kappa_-^2+2\nu_-^2-K_+(a_-)\right).\nonumber
\ea\eeqn
Hence,
$$
\tilde D(\pm a_-)\not=0\,\,\, \Longleftrightarrow\,\,\,
\mu_0\not=\kappa^2_-(m_0-m_+)+4\nu_-^2\left(m_0-\frac{m_-+m_+}{2}\right)+m_+K_+(a_-).
$$
If condition ${\bf C}$ holds, then (\ref{lemmaA1}) is true
with $\omega_0=\pm a_-$,
 $C_1=(\tilde D(\pm a_-))^{-1}$.
If condition ${\bf C}_0$~{\bf(i)} holds, then
(\ref{lemmaA2}) is true.
\medskip

{\bf(3.2)} In the case when $a_-=a_+$,
we apply (\ref{a9}) to $e^{i\theta_\pm(\omega)}$ and obtain
$$
\tilde D(\omega)
=\tilde D(\pm a_-)-i(\sqrt{m_-\gamma_-}+\sqrt{m_+\gamma_+})(a_-^2-\omega^2)^{1/2}+O(|\omega\mp a_-|),
\,\,\,\omega\to\pm a_-,
$$
where $\tilde D(\pm a_-)=\mu_0-m_0a_-^2+2\gamma_-+2\gamma_+$. Hence,
$$
\tilde D(\pm a_-)\not=0\quad \Longleftrightarrow\quad
\mu_0\not=m_0a_-^2- 2\gamma_--2\gamma_+.
$$
Condition ${\bf C}$ implies (\ref{lemmaA1})  with $\omega_0=\pm a_-$.
If condition ${\bf C}_0$~{\bf(iii)} holds, then
 (\ref{lemmaA2}) is true with $\omega_0=\pm a_-$.
\medskip

{\bf(3.3)} If $a_-\in(\kappa_+,a_+)$, then
we apply (\ref{a9}) to $e^{i\theta_-(\omega)}$ and obtain
$$
\tilde D(\omega)
=\tilde D( a_-\pm i0)-i\sqrt{m_-\gamma_-}(a_-^2-\omega^2)^{1/2}+O(|\omega- a_-|),
\,\,\,\omega\to a_-\pm i0,
$$
where
$\tilde D(a_-\pm i0)=\mu_0-m_0a_-^2+2\gamma_-+\gamma_+
\left(1- e^{\pm i\theta_+( a_-+ i0)}\right)$
with  $\Im\theta_+( a_-+ i0)=0$.
$\Im\tilde D(a_-\pm i0)\not=0$, since
$$
\Im\tilde D(a_-\pm i0)
=\mp\frac{m_+}{2}\left\{\ba{lll}
\sqrt{a_-^2-\kappa^2_+}\sqrt{a_+^2-a_-^2},&\mbox{if }\, \kappa_+>0,\\
 2a_-\sqrt{\nu_+^2-\nu_-^2},&\mbox{if }\, \kappa_+=\kappa_-=0.
\ea\right.
$$
Hence, $\tilde D(a_-+i0)\not=0$ and (\ref{lemmaA1}) follows.
Similarly, for $\omega\to -a_-\pm i0$.
\medskip

{\bf(3.5)} If $a_-<\kappa_+$, then
$e^{i\theta_+(\omega)}$ is an analytic function
in a small neighborhood of the points $\omega_0=\pm a_-$.
Applying the representation (\ref{a9}) to $e^{i\theta_-(\omega)}$, we obtain
$$
\tilde D(\omega)=
\tilde D(\pm a_-)-i\sqrt{m_-\gamma_-}(a_-^2-\omega^2)^{1/2}+O(|\omega\mp a_-|),
\,\,\,\omega\to\pm a_-,\quad\omega\in\mathbb{C}_+,
$$
where (see (\ref{a}))
\beqn
\tilde D(a_-)&=&\mu_0-m_0a_-^2+2\gamma_-+\gamma_+(1-e^{i\theta_+(a_-)})
\nonumber\\
&=&\mu_0-\kappa_-^2(m_0-m_+)-4\nu_-^2\left(m_0-\frac{m_-+m_+}{2}\right)-m_+K_0(a_-).\nonumber
\eeqn
Hence,
$$
\tilde D(\pm a_-)\not=0\,\,\, \Longleftrightarrow\,\,\,
\mu_0\not=\kappa_-^2(m_0-m_+)+4\nu_-^2\left(m_0-\frac{m_-+m_+}{2}\right)+m_+K_0(a_-).
$$
Condition~{\bf C} implies (\ref{lemmaA1})  with $\omega_0=\pm a_-$.
If condition ${\bf C}_0$~{\bf(vi)} holds, then
(\ref{lemmaA2}) is true with $\omega_0=\pm a_-$.
\medskip

{\bf(4)} Let $\omega_0=\pm a_+$. There are the following cases:
{\bf(4.1)} $a_+<a_-$; {\bf(4.2)} $a_+= a_-$ (see  case {\bf(3.2)}); {\bf(4.3)} $a_+>a_-$.
\medskip

{\bf(4.1)} If $a_+<a_-$, then $a_+\in\Lambda_-\setminus \Lambda_-^0$.
Similarly to  case {\bf(3.3)},  we obtain
$$
\tilde D(\omega)
=\tilde D(a_+\pm i0)-i\sqrt{m_-\gamma_-}(a_+^2-\omega^2)^{1/2}+O(|\omega- a_+|),
\,\,\,\omega\to a_+\pm i0.
$$
$\tilde D(a_+\pm i0)=\mu_0-m_0a_+^2+2\gamma_++\gamma_-
\left(1- e^{\pm i\theta_-( a_++ i0)}\right)$
with  $\Im\theta_-( a_++ i0)=0$. Since
$$
\Im\tilde D(a_+\pm i0)
=\mp\frac{m_-}{2}\left\{\ba{lll}
\sqrt{a_+^2-\kappa^2_-}\sqrt{a_-^2-a_+^2},&\mbox{if }\, \kappa_->0\\
 2a_+\sqrt{\nu_-^2-\nu_+^2},&\mbox{if }\, \kappa_-=0,
\ea\right.
$$
$\Im\tilde D(a_+\pm i0)\not=0$.
Hence, $\tilde D(a_++i0)\not=0$ and (\ref{lemmaA1}) follows.
Similarly, for $\omega\to -a_+\pm i0$.
\medskip

{\bf(4.3)} If $a_+>a_-$, then
$e^{i\theta_-(\omega)}$ is an analytic function
in a small neighborhood of the points $\omega_0=\pm a_+$.
Using (\ref{a9}), we have
$$
\tilde D(\omega)=
\tilde D(\pm a_+)-i\sqrt{m_-\gamma_-}(a_+^2-\omega^2)^{1/2}+O(|\omega\mp a_+|),
\,\,\,\omega\to\pm a_+.
$$
where by (\ref{b}),
\beqn
\tilde D(a_+)&=&
\mu_0-m_0a_+^2+2\gamma_++\gamma_-(1+e^{-\Im\theta_-(a_+)})
\nonumber\\
&=&\mu_0-\kappa_+^2(m_0-m_-)-4\nu_+^2\left(m_0-\frac{m_-+m_+}{2}\right)
-m_-K_-(a_+).\nonumber
\eeqn
Hence,
$$
\tilde D(\pm a_+)\not=0\,\,\, \Longleftrightarrow\,\,\,
\mu_0\not=\kappa_+^2(m_0-m_-)+4\nu_+^2\left(m_0-\frac{m_-+m_+}{2}\right)
+m_-K_-(a_+).
$$
Condition ${\bf C}$ implies (\ref{lemmaA1})  with $\omega_0=\pm a_+$.
If condition ${\bf C}_0$~{\bf(ii)} holds, then
 (\ref{lemmaA2}) is true with $\omega_0=\pm a_+$.
Lemma~\ref{lemmaA} is proved.
\bo

\subsection{Proof of Theorem~\ref{l3.1}}
By Lemma~\ref{l2.A}, we can vary the contour of integration in (\ref{N}) as follows:
\be\label{b.10}
{\cal N}(t)=-\frac1{2\pi}\int_{|\omega|=R}
e^{-i\omega t} \tilde {\cal N}(\omega)\,d\omega,\quad t>0.
\ee
where the number $R$ is chosen enough large, $R>\max\{a_-,a_+\}$.
By Lemmas~\ref{l2.A} and \ref{l8.2}, we rewrite ${\cal N}(t)$ in the form
$$
{\cal N}(t)=-\frac1{2\pi}\int_{\Lambda_{\ve}}
e^{-i\omega t} \tilde {\cal N}(\omega)\,d\omega,\quad t>0,
$$
where $\varepsilon\in(0,\delta)$,
the contour $\Lambda_{\ve}$ surrounds the segments of $\Lambda$
and belongs to $\ve$-neighborhood of  $\Lambda$
(the contour $\Lambda_{\ve}$ is  oriented anticlockwise).
Passing to a limit as $\ve\to0$, we obtain
 \beqn
 {\cal N}(t)&=&\frac1{2\pi}\int_{\Lambda}
 e^{-i\omega t}\left(\tilde {\cal N}(\omega+i0)-\tilde {\cal N}(\omega-i0)\right)\,d\omega
 \nonumber\\
 &=&
  \sum\limits_{\pm}\sum\limits_{j=1}^2\frac{1}{2\pi}\int_{\Lambda}
 e^{-i\omega t} P_j^{\pm}(\omega)\,d\omega+o(t^{-K}),\quad t\to+\infty,
 \quad \mbox{with any  }\, K>0.  \nonumber
\eeqn
Here $P_j^{\pm}(\omega):=\zeta_j^\pm(\omega)(\tilde {\cal N}(\omega+i0)-\tilde {\cal N}(\omega-i0))$,
 $\zeta_j^\pm(\omega)$ are smooth functions such that
$\sum\limits_{\pm,j}\zeta_j^\pm(\omega)=1$, $\omega\in\R$,
$\supp\zeta_1^\pm\subset {\cal O}(\pm \kappa_\pm)$,
$\supp\zeta_2^\pm\subset {\cal O}(\pm a_\pm)$
(${\cal O}(b)$ denotes a neighborhood of the point $\omega=b$).
In the case  $\kappa_\pm=0$, instead of $\zeta_1^\pm$
($P_1^\pm$) we introduce  the function $\zeta_1$ (respectively, $P_1$)
with $\supp\zeta_1\subset {\cal O}(0)$.
Then Lemma~\ref{lemmaA} implies the bound~(\ref{NN}) with  $k=0$.
Here we use the following estimate
$$
\Big|\int_{\R}\zeta(\omega) e^{-i\omega t}
(a^2-\omega^2)^{j/2}\,d\omega\Big|\le C(1+t)^{-1-j/2}\quad
\mbox{при }\,t\to+\infty,
$$
where $j=\pm1$ ($j=-1$, if condition~${\bf C}_0$ holds, $j=1$, if condition~${\bf C}$ holds),
 $\zeta(\omega)$  is a smooth function,
and $\zeta(\omega)=1$ for $|\omega-a|\le\delta$ with some $\delta>0$.
The bound (\ref{NN}) with $k=1,2$ can be proved by a similar way.
\bo

\subsection{Resonance cases:  $N=0$}\label{Sec3.6}

 The cases of the constants when conditions~{\bf C} and $\mathbf{C}_0$ are not fulfilled we call
by {\em resonance cases}.
In these cases,   ${\cal N}(t)=C_1+C_2\sin(a t+b)+O(t^{-1/2})$ as $t\to\infty$
 with some constants $C_1,C_2,a,b$.
In the particular case {\bf(P2)}, the resonance cases were studied
 in \cite{rjmp-2018}.
 Let us consider the particular case~{\bf (P1)}.
  Then, the resonance cases are the following three cases:
\begin{itemize}
  \item[] case $(R1)$: $\kappa_0=\kappa=0$;
  \item[] case $(R2)$: $(\kappa_0,\kappa)\not=(0,0)$ and $a_0>a$;
  \item[] case $(R3)$: $\kappa_0<\kappa$.
\end{itemize}
\medskip
Resonance  cases can be rewritten in the terms of restrictions on $m_0,m,\gamma,\mu_0,\mu$ as follows
\begin{itemize}
  \item{}  $\mu_0=\mu=0$
  \item{}  $(\mu_0,\mu)\not=(0,0)$ and $m_0<m$
  \item{}  $\mu_0<\mu$
\end{itemize}
\medskip

We construct the solutions $u(\cdot,t)$ which do not satisfy the bound (\ref{0.3}) and (\ref{2.31'}).
  In the case (R1), $\tilde {\cal N}(\omega)$ has a simple pole at zero,
  $$
  \tilde {\cal N}(\omega)=\frac{i}{2\omega \sqrt{\gamma\,m }}+
  \frac{m-m_0}{4\gamma\, m}+\dots\quad\mbox{as }\,\,\omega\to0.
    $$
Hence,
 ${\cal N}(t)=(2\sqrt{\gamma\,m })^{-1}+O(t^{-1/2})$ as $t\to\infty$.
Suppose that $Y_0(n)\equiv0$ for $n\not=0$.
Then, $Y_0\in{\cal H}_{\alpha}$ for any $\alpha$ and $z(\cdot,t)\equiv0$.
Using (\ref{2.1}) and (\ref{2.34}), we obtain
$$
u(0,t)=r(0,t)=\frac{v_0(0)}{2\sqrt{\gamma\,m}}+O(t^{-1/2})\quad\mbox{as }\,\, t\to\infty.
$$
Hence, if $v_0(0)\not=0$, then
the constructed solution $u(n,t)$ does not satisfy the bound (\ref{2.31'}).
\medskip

In the cases (R2) and (R3),
 there exists a point $\omega_0\in\R\setminus\Lambda$ such that
$\tilde {\cal N}(\omega)$ has simple poles at the points $\omega=\pm\omega_0$
with $\omega_0\in\R\setminus\Lambda$.
Hence, ${\cal N}(t)=C\sin(\omega_0 t)+O(t^{-3/2})$ as $t\to\infty$.
Note that $\Im\theta(\omega_0)>0$, $\Re\theta(\omega_0)=\pm\pi$.
Therefore, the function of the form
$$
u(n,t)=e^{i\theta(\omega_0)|n|}\sin(\omega_0 t),\quad t\ge0
$$
is a solution of the system.
However, this solution does not satisfy the bound~(\ref{2.31'}).

\setcounter{equation}{0}
\section{Appendix B: Case $N\ge1$}\label{appendixB}
\subsection{$\tilde D(\omega)$ for $\omega\in\CC_\pm$}
We prove the following lemma  for any $N\ge0$ without the additional condition (\ref{2.30}).
\begin{lemma}\label{l2.A}
(i) $\tilde {\cal N}(\omega)$
is meromorphic-valued matrix for $\omega\in \mathbb{C}\setminus\Lambda$.

(ii) $|\tilde {\cal N}(\omega)|=O(|\omega|^{-2})$ as $|\omega|\to\infty$.

(iii) $\textrm{det }\tilde D(\omega)\not=0$  for all $\omega\in \mathbb{C}_\pm$.
\end{lemma}
{\bf Proof}\,
The first assertion of the lemma  follows from the formula (\ref{3.21})
and the analyticity of
$\tilde D(\omega)$  for $\omega\in \mathbb{C}\setminus \Lambda$.
By (\ref{32}) and the condition $\Im\theta_\pm(\omega)>0$, we have
\be\label{7.0}
\left| e^{i\theta_\pm(\omega)}\right|\le C|\omega|^{-2}\quad \mbox{as }\,\,|\omega|\to\infty.
\ee
Then the second assertion follows from (\ref{3.21}) and (\ref{7.0}).
The  third assertion follows from the energy bound~(\ref{H-1}). Indeed,
let us assume that $\textrm{det }\tilde D(\omega_0)=0$ for some $\omega_0\in \mathbb{C}_+$.
Then, there exists a nonzero vector $\xi=(\xi_0,\dots,\xi_N)$ such that
\begin{equation}\label{a1}
\tilde D(\omega_0)\xi=0.
\end{equation}
Introduce a function $u_*(n,t)$, $n\in\Z$, $t\ge0$, as
$$
u_*(n,t)=\left\{\ba{lll}
e^{-i\omega_0 t}e^{-i\theta_-(\omega_0)n}\xi_0,&n\le0,\\
e^{-i\omega_0 t}\xi_1,&n=1,\\
\dots&\dots\\
e^{-i\omega_0 t}\xi_{N-1},&n=N-1,\\
e^{-i\omega_0 t}e^{i\theta_+(\omega_0)(n-N)}\xi_N,&n\ge N.
\ea\right.
$$
 Using (\ref{32}) and (\ref{a1}),
 it is easy to check that $Y_*(t)=\left(u_*(\cdot,t),v_*(\cdot,t)\right)$,
 where $v_*(n,t)=m_n\dot u_*(n,t)$, is a solution of the problem (\ref{1})
with the initial data
$$
Y_*(n)=\left\{\ba{lll}
e^{-i\theta_-(\omega_0)n}\xi_0(1,-im_-\omega_0),&n\le0,\\
\xi_1(1,-im_0\omega_0),&n=1,\\
\dots&\dots\\
\xi_{N-1}(1,-im_N\omega_0),&n=N-1,\\
e^{i\theta_+(\omega_0)(n-N)}\xi_N(1,-im_+\omega_0),&n\ge N.
\ea\right.
$$
Therefore, the Hamiltonian  (see (\ref{H1})) is
$$
{\bf H}(Y_*(t))=e^{2t\,\Im\omega_0}{\bf H}(Y_*(\cdot))
\quad \mbox{for any } t>0,
\quad \mbox{where }\,\, {\bf H}(Y_*)>0.
$$
Since $\Im\omega_0>0$ and $Y_*\in {\cal H}_0$,
  this exponential growth  contradicts the energy estimate (\ref{H-1}).
Hence, $\textrm{det }\tilde D(\omega)\not=0$ for any $\omega\in \mathbb{C}_+$.
Since $\overline{\theta_\pm(\omega)}=-\theta_\pm(\bar\omega)$ for $\omega\in \mathbb{C}\setminus \Lambda$,
then $\overline{\tilde D(\omega)}=\tilde D(\bar\omega)$.
Therefore, $\textrm{det }\tilde D(\omega)\not=0$ for any $\omega\in \mathbb{C}_-$.
\bo

\subsection{$\tilde D(\omega)$ for $\omega\in\Lambda\setminus \Lambda^0$}
For simplicity of the further  calculations, we impose condition~(\ref{2.30}).
Then, the system~(\ref{1-})--(\ref{1+}) is of a form
$$
\left\{\ba{lll}
m\ddot u(n,t)=\left(\gamma\Delta_L-\mu\right)u(n,t)& \mbox{for }\,n\le -1,\quad n\ge N+1,\\
m_n\ddot u(n,t)=\gamma_n\nabla_Lu(n,t)-\gamma_{n-1}\nabla_Lu(n-1,t)-\mu_nu(n,t)&\mbox{for }\,n=0,1,\dots,N,
\ea\right.
$$
with $\gamma_N=\gamma_{-1}\equiv\gamma$.
Set
$\nu:=\nu_\pm$, $\kappa:=\kappa_\pm$, $a:=a_\pm$, where
$\nu^2=\gamma/m$, $\kappa^2=\mu/m$, $a^2=(\mu+4\gamma)/m$,
$\theta(\omega):=\theta_\pm(\omega)$,
$\Lambda=[-a,-\kappa]\cup[\kappa,a]$, $\Lambda^0=\{\pm\kappa,\pm a\}$.
\begin{lemma}\label{l5.2}
$\textrm{det }\tilde D(\omega\pm i0)\not=0$ for $\omega\in\Lambda\setminus \Lambda^0$.
\end{lemma}
{\bf Proof }\,
For $N=0$, we prove this result in Appendix~A.
Now we assume that $N\ge1$.
For $\omega\in\Lambda\setminus \Lambda^0$, $\Re\theta(\omega+i0)\in(-\pi,0)\cup(0,\pi)$
and $\Im\theta(\omega+i0)=0$.
Moreover, ${\rm sign}(\sin\theta(\omega+i0))={\rm sign}\,\omega$.
 Write
 $$
 y:=\Im\tilde D_{00}(\omega+i0)=\Im\tilde D_{NN}(\omega+i0)=-\gamma\sin\theta(\omega+i0)\not=0;
 $$
$$
r_0:=\Re\tilde D_{00}(\omega+i0),\quad r_N:=\Re\tilde D_{NN}(\omega+i0),
$$
$$
d_n\equiv d_n(\omega):=\tilde D_{nn}(\omega)\in\R,\quad n=1,\dots,N-1.
$$

Assume the contrary, that
 $\textrm{det }\tilde D(\omega+ i0)=0$. Then,
by (\ref{4.10}),
$$
\textrm{det }\tilde D(\omega+ i0)=\alpha_N(\omega+i0)=(r_N+iy)\alpha_{N-1}-\gamma_{N-1}^2\alpha_{N-2},
$$
where $\alpha_{-2}=0$, $\alpha_{-1}=1$ and
 $\alpha_n\equiv \alpha_n(\omega+i0)$ for $n=0,\dots,N$,
 $\alpha_n(\omega)$ for $\omega\in\CC$  are defined in (\ref{alphai}).
Hence,
\be\label{4.7}
\left\{\ba{l}
r_N\,\Re\alpha_{N-1}-y\,\Im\alpha_{N-1}-\gamma_{N-1}^2\Re\alpha_{N-2}=0\\
y\,\Re\alpha_{N-1}+r_N\,\Im\alpha_{N-1}-\gamma_{N-1}^2\Im\alpha_{N-2}=0
\ea
\right.
\ee
At first, we prove that equalities in (\ref{4.7}) is impossible in the case $N=1$.
Indeed, in this case,
 $\alpha_{N-1}\equiv \alpha_0=r_0+iy$, $\alpha_{N-2}\equiv \alpha_{-1}=1$,
 and the system~(\ref{4.7}) becomes
\beqn
r_0r_1-y^2-\gamma_0^2=0\label{4.5}\\
yr_0+yr_1=0\label{4.6}
\eeqn
By (\ref{4.6}), $r_0=-r_1$, since $y\not=0$. Then,  using~(\ref{4.5}), we have
$-r_0^2-y^2-\gamma_0^2=0$
what is impossible. Therefore, $\textrm{det }\tilde D(\omega+i0)\not=0$
for any $\omega\in\Lambda\setminus\Lambda^0$ in the case $N=1$.

Now we prove that equalities in (\ref{4.7}) is impossible if $N\ge2$.
 For $N\ge2$, introduce the following determinants $\Delta_k^j\equiv\Delta_k^j(\omega)$ by the rule
\begin{equation}\label{deltai}
\Delta_k^j(\omega)={\rm det}\left(
  \begin{array}{cccccc}
    d_{j}&-\gamma_j&0&\ldots&0 \\
    -\gamma_j& d_{j+1}&  -\gamma_{j+1}&\ddots&0\\
    0&-\gamma_{j+1}&\ddots&\ddots&0\\
    \vdots&\ddots&\ddots&d_{k-1}&-\gamma_{k-1}\\
    0 &\ldots&0&-\gamma_{k-1}&d_{k} \\
  \end{array}
\right),\,\,
\ba{ll}
j\le k\\k=1,\dots,N\!-\!1\\
N\ge2.\ea
\end{equation}
The matrices in (\ref{deltai}) are real-valued symmetric and tridiagonal
(i.e., normal Jakobi matrices).
Moreover, for fixed $j$, the minors $\{\tilde \alpha_p\equiv \Delta_k^j|_{k=j+p}\,, p=0,1,\dots\}$
satisfy the recurrence relation (cf (\ref{4.10}))
\be\label{4.10'}
\tilde \alpha_p
=d_{j+p}\,\tilde\alpha_{p-1}-\gamma_{j+p-1}^{2}\tilde\alpha_{p-2},\quad p=0,1,\dots,
\ee
with the initial conditions
$\tilde\alpha_{-2}=0$ and $\tilde\alpha_{-1}=1$. In turn,
for fixed $k$, the minors $\{\tilde \beta_p\equiv \Delta_k^j|_{j=p}\,, p=k,k-1,\dots\}$
 satisfy the three-term recurrence (cf (\ref{4.11}))
\be\label{4.11'}
\tilde\beta_p=d_{p}\tilde\beta_{p+1}-\gamma_{p}^{2}\tilde\beta_{p+2},\quad p=k,k-1,\dots,
\ee
with the initial conditions
$\tilde\beta_{k+1}=1$, $\tilde\beta_{k+2}=0$.
We return to the proof of Lemma~\ref{l5.2} in the case $N\ge2$.
Applying notation~(\ref{deltai}), we have
 $\alpha_1=(r_0+iy)d_1-\gamma_0^2$,
$\alpha_k=(r_0+iy)\Delta_k^1-\gamma_0^2\Delta_k^2$, $k=2,\dots,N-1$,
$N\ge2$.
Since $y\not=0$ and $\Delta_k^j\in\R$ for $\omega\in\R$,
the system (\ref{4.7}) becomes
\beqn
r_N\Re\alpha_{N-1}-y^2\Delta_{N-1}^1-\gamma_{N-1}^2\Re\alpha_{N-2}=0
\label{4.8}\\
\Re\alpha_{N-1}+r_N\Delta_{N-1}^1-\gamma_{N-1}^2\Delta^1_{N-2}=0
\label{4.9}
\eeqn
where $\Re\alpha_{k}=r_0\Delta_{k}^1-\gamma_{0}^2\Delta^2_{k}$,
$\Delta_{0}^1\equiv1$ and $\Delta_{0}^2\equiv0$.
\smallskip

Assume that $\Delta_{N-1}^1=0$. Then, $\Delta^2_{N-1}\cdot\Delta^1_{N-2}>0$.
This inequality follows from two facts for determinants $\Delta_{k}^j$:
(1)~if $\Delta_{N-1}^1=0$, then  $\Delta^2_{N-1}\cdot\Delta^1_{N-2}\not=0$
by the properties of the minors of Jakobi matrices or by the relations~(\ref{4.10'}) and (\ref{4.11'});
(2)~nonzero major minors of same order (as $\Delta^2_{N-1}$ and $\Delta^1_{N-2}$) have the same sign
(see, e.g., \cite[p.31, p.83 in Russian edition]{GK}).
However, this inequality contradicts to (\ref{4.9}), because
by (\ref{4.9}) and $\Delta_{N-1}^1=0$, one obtains
$-\gamma_{0}^2\Delta^2_{N-1}-\gamma_{N-1}^2\Delta^1_{N-2}=0$
and then, $\textrm{sign}\Delta^2_{N-1}=-\textrm{sign}\Delta^1_{N-2}$.
\smallskip

If $\Delta_{N-1}^1\not=0$, then we express $r_N$ from Eqn~(\ref{4.9}) and
substitute in (\ref{4.8}).
For $N=2$, if $\Delta_{N-1}^1\equiv \Delta_{1}^1=d_1\not=0$, then
$$
r_2=\frac{\gamma_1^2\Delta_0^1-\Re\alpha_1}{\Delta_1^1}=
\frac{\gamma_1^2-(x_0d_1-\gamma_0^2)}{d_1}
$$
and  Eqn~(\ref{4.9}) becomes
$$
\frac{-(x_0d_1-\gamma_0^2)^2-y^2d_1^2-\gamma_0^2\gamma_1^2}{d_1}=0,
$$
what is impossible since $\gamma_0\gamma_1\not=0$.
For $N\ge3$, we use the following equality
$$
\Delta_{N-1}^2\cdot\Delta_{N-2}^1-\Delta_{N-2}^2\cdot\Delta_{N-1}^1
=\gamma_1^2\cdot\ldots\cdot\gamma_{N-2}^2,
\,\,\,\mbox{where }\,N\ge3,\,\,\,\Delta_{1}^2\equiv1,
$$
which can be proved by induction.
Therefore, Eqn~(\ref{4.8}) writes
$$
\frac{-\left(\Re\alpha_{N-1}\right)^2-y^2\left(\Delta_{N-1}^1\right)^2-
\gamma_0^2\gamma_1^2\cdot\ldots\cdot\gamma^2_{N-1}}{\Delta_{N-1}^1}=0
$$
what is impossible.
Thus, $\textrm{det }\tilde D(\omega+i0)\not=0$
for any $\omega\in\Lambda\setminus\Lambda^0$.
Since
$\tilde D(\omega-i0)=\overline{\tilde D(\omega+i0)}$
for $\omega\in\Lambda\setminus\Lambda^0$,
$\textrm{det }\tilde D(\omega-i0)\not=0$ for $\omega\in\Lambda\setminus\Lambda^0$.
\bo

\subsection{$\tilde D(\omega)$ for $\omega\in\R\setminus \Lambda$}

For $\omega\in\R\setminus \Lambda$, the matrix $\tilde D(\omega)$ is symmetric and real-valued.
As before, we write $d_n(\omega)=\tilde D_{nn}(\omega)$.
We consider separately two cases of values of $\omega$:
$|\omega|>a$ and $|\omega|<\kappa$ (if $\kappa\not=0$).
For $|\omega|>a$, the following result holds.
\begin{lemma}\label{l5.3}
Write $\R_a:=\{\omega\in\R:|\omega|>a\}$. The following assertions are equivalent.
\begin{description}
\item[(A1)]
${\rm det }\,\tilde D(\omega)\not=0$ for any $\omega\in\R_a$.

 \item[(A2)]
 $(-1)^N{\rm det }\,\tilde D(\omega)<0$ for any $\omega\in\R_a$.

\item[(A3)]
For every $n=0,1,\dots,N$,
$\alpha_n(\omega)(-1)^n<0$ for any $\omega\in\R_a$, i.e.,
the symmetric matrix $\tilde D(\omega)$ is negative definite for $\omega\in\R_a$.

\item[(A4)] For every $n=0,1,\dots,N$,
$\beta_n(\omega)(-1)^{N+n}<0$ for any $\omega\in\R_a$.

\item[(A5)] For every $n=0,1,\dots,N$,
$c_n(\omega)<0$ for any $\omega\in\R_a$.

\item[(A6)] $\alpha_n(a)(-1)^n<0$ for $n=0,1,\dots,N-1$,
 and $\alpha_N(a)(-1)^N\le 0$.

\item[(A7)] $c_n(a)<0$ for $n=1,\dots,N-1$, and $c_N(a)\le 0$.
\end{description}
\end{lemma}
{\bf Proof}\, For $\omega\in\R_a$, $\tilde D(\omega)$ is a real-valued symmetric (Jakobi)
 matrix with diagonal terms of a form
 $$
 \ba{lll}
 d_{0}(\omega)=\mu_0-m_0\omega^2+\gamma(1+e^{-\Im\theta(\omega)})+\gamma_0,\\
  d_{n}(\omega)=\mu_n-m_n\omega^2+\gamma_n+\gamma_{n-1},
 \quad n=0,\dots N-1,\\
  d_{N}(\omega)=\mu_N-m_N\omega^2+\gamma(1+e^{-\Im\theta(\omega)})+\gamma_{N-1}.
\ea
$$
Note that for $\omega\in\R_a$,
 \be\label{4.22}
d_{n}(\omega_1)< d_{n}(\omega_2)\,\,\,\mbox{for }\,
|\omega_1|>|\omega_2|\quad
\mbox{and }\,\, d_{n}(\omega)\to-\infty\quad\mbox{as }\,\, \omega\to\pm\infty.
\ee
It follows from (\ref{4.22}) and (\ref{4.18}) that
all $c_n(\omega)$ are even, strictly decrease for $\omega>a$ and $c_n(\omega)\to-\infty$
for $\omega\to\pm\infty$.
Furthermore, (\ref{4.22}) (or (\ref{4.17})) implies that
\be\label{4.23}
(-1)^N{\rm det }\,\tilde D(\omega)\to-\infty\,\,\,\,\mbox{as }\,\, \omega\to\pm\infty.
\ee
Hence, assertion~(A1) is equivalent to (A2).
Furthermore, (A7) $\Longrightarrow$ (A5).
By formula~(\ref{4.17}), (A5) implies (A2).
Evidently, (A3) $\Longrightarrow$ (A2). 
Assertions~(A3), (A4) and (A5) are equivalent by Remark~\ref{r5.6},
and (A6) $\Longleftrightarrow$ (A7). 
Therefore, (A6) $\Longrightarrow$~(A2).
\smallskip

It remains to prove that (A2) implies (A3) and (A6).
Assume, for simplicity, that $N=1$. Let assertion~(A2) hold.
Then, $\alpha_0(a)\equiv d_0(a)\le0$.
Indeed, if $d_0(a)>0$, then, by (\ref{4.22}),
there is a point $\omega_1>a$ such that
$d_0(\omega_1)=0$.  Hence, ${\rm det }\,\tilde D(\omega_1)=-\gamma^2_0<0$.
This contradicts to (A2).
Similarly, we can check that (A2) implies that $d_1(a)\le0$.
Hence, $d_0(\omega)<0$ and $d_1(\omega)<0$ for all  $\omega\in\R_a$, by (\ref{4.22}),
and  (A3) holds. Moreover,
$\alpha'_1(\omega)\equiv({\rm det }\,\tilde D(\omega))'>0$ for $\omega>a$. Therefore,
 $\alpha_1(a)\ge0$.
If $\alpha_0(a)=0$, then $\alpha_1(a)=-\gamma_0^2<0$ what is impossible.
Hence, $\alpha_0(a)>0$ and (A6) is true.
In the case $N\ge 2$, the proof of implications (A2)~$\Longrightarrow$~(A3)
and (A2)~$\Longrightarrow$~(A6) is similar
and is based on two facts:
\be\label{apb-0}
1) \quad  \alpha_n(\omega)(-1)^n\to-\infty\quad \mbox{as }\,\,\omega\to\pm\infty,
\ee
and 2) if for some point $\omega_0$, $\alpha_n(\omega_0)=0$, then
$\alpha_{n-1}(\omega_0)\alpha_{n+1}(\omega_0)<0$
 by virtue of (\ref{4.10}).
\smallskip

Now we prove these implications for $N=2,3$.
Assume that $\alpha_0(a)\equiv d_0(a)>0$.
It follows from (\ref{4.22}) that there exists a point $\omega_1>a$ such that
$\alpha_0(\omega_1)=0$ and
\be\label{apb-1}
\alpha_0(\omega)<0\quad \mbox{for all }\,\, \omega>\omega_1.
\ee
Then, $\alpha_1(\omega_1)=-\gamma_0^2<0$.
Therefore, by (\ref{apb-0}), there is a point $\omega_2>\omega_1$ such that
\be\label{apb-2}
\alpha_1(\omega_2)=0.
\ee
Hence,
\be\label{apb-3}
\alpha_2(\omega_2)=-\gamma_1^2\alpha_0(\omega_2)>0,
\ee
by (\ref{apb-1}).
If $N=2$, then (\ref{apb-3}) contradicts (A2). Hence, $\alpha_0(a)\le0$
and
\be\label{apb-4}
\alpha_0(\omega)<0\quad\mbox{ for any }\,\, |\omega|>a,
\ee
by (\ref{4.22}).
Moreover, from reasonings above we see that
$\alpha_1(\omega)>0$ for all $\omega\in\R_a$ and (A3) is true.
If $\alpha_0(a)=0$, then $\alpha_1(a)=-\gamma_0^2<0$ and there is a point $\omega_1>a$
such that $\alpha_1(\omega_1)=0$ that is impossible by (A3).
Hence, $\alpha_0(a)<0$. Also, $\alpha_1(a)\not=0$, since if $\alpha_1(a)=0$, then
$\alpha_2(a)=-\alpha_0^2\alpha_0(a)>0$, what contradicts (A3).
Therefore, (A6) is true.
If $N=3$, then by (\ref{apb-0}) and (\ref{apb-3}),  there is a point $\omega_3>\omega_2$
such that $\alpha_2(\omega_3)=0$. Hence, $\alpha_1(\omega_3)\alpha_3(\omega_3)<0$
by (\ref{4.10}).
We can choose a point $\omega_2$ in (\ref{apb-2}) such that
$\alpha_1(\omega)>0$ for any $\omega>\omega_2$.
Hence, $\alpha_3(\omega_3)<0$. But this inequality contradicts (A2) with $N=3$.
Hence, (\ref{apb-4}) is valid.
Moreover, using  reasonings above we obtain that $\alpha_1(\omega)<0$ and $\alpha_2(\omega)>0$
for any $\omega\in\R_a$ and (A3) is true.

Now we check (A6). Since $c_n(\omega)<0$
for any $n$ and $\omega\in\R_a$, then $(-1)^n\alpha_n(\omega)<0$ for any $\omega\in\R_a$
by virtue to Remark~\ref{r5.6}~(2). Hence,
\be\label{apb-5}
(-1)^n\alpha_n(a)\le0\quad \mbox{for any }\,\,n.
\ee
It remains to prove that $\alpha_n(a)\not=0$ for $n\not=N$.
We check this  fact  by induction.
Indeed, if $\alpha_0(a)=0$, then $\alpha_1(a)<0$, what is impossible by (\ref{apb-5}).
Hence, $\alpha_0(a)<0$.
Assume that $(-1)^k\alpha_k(a)<0$ for $k\le n-1$.
If $\alpha_n(a)=0$, then $(-1)^{n-1}\alpha_{n-1}(a)(-1)^{n+1}\alpha_{n+1}(a)<0$.
Hence, $(-1)^{n+1}\alpha_{n+1}(a)>0$ what contradicts (\ref{apb-5}).
\bo
\medskip

Let $\kappa\not=0$.
At first, we prove the following auxiliary lemma.
\begin{lemma}\label{l5.5}
If $\kappa\not=0$, then $c_n(0)>0$ and $\alpha_n(0)>0$ for any $n=0,1,\dots,N$.
\end{lemma}
{\bf Proof}\, At first note that
$$
K:=1-e^{-i\theta(0)}=4\left(\frac{\kappa}{\nu}+\sqrt{1+\frac{\kappa^2}{\nu^2}}\right)^{-2}<1.
$$
Then, if $N=0$, then by (\ref{D-N0}), we have
$$
\tilde D(0)\equiv c_0(0)\equiv \alpha_0(0)=\mu_0+2\gamma K>0.
$$
If $N\ge1$, then
$$
\left\{\ba{lllll}
d_0(0)&=&\mu_0+\gamma K+\gamma_0&>&\gamma_0,\\
d_n(0)&=&\mu_n+\gamma_n+\gamma_{n-1}&\ge&\gamma_n+\gamma_{n-1}\quad\mbox{ for }\, n=1,\dots, N-1,\\
d_N(0)&=&\mu_N+\gamma K+\gamma_{N-1}&\ge&\gamma K+\gamma_{N-1}.
\ea\right.
$$
Hence, using  (\ref{4.18}), we have
$c_n(0)>\gamma_n$ for $n=0,1,\dots,N-1$, and $c_N(0)> \gamma K>0$.
Since $\alpha_0(0)=c_0(0)$ and $\alpha_n(0)=\alpha_{n-1}(0)c_n(0)$, then
$\alpha_n(0)>\gamma_0\cdot\dots\cdot\gamma_n$ for  $n=0,\dots,N-1$ and
$\alpha_N(0)>\gamma_0\cdot\dots\cdot\gamma_{N-1}\gamma K>0$.
By Sylvester's criterion, the symmetric matrix $\tilde D(0)$ is positive definite.
\bo
\begin{lemma}\label{l5.4}
 Write $\R_\kappa:=\{\omega\in\R:|\omega|<\kappa\}$.
Then the following assertions are equivalent.

(A1) ${\rm det }\,\tilde D(\omega)\not=0$ for any $\omega\in\R_\kappa$.

(A2) ${\rm det }\,\tilde D(\omega)>0$ for any $\omega\in\R_\kappa$.

(A3) For every $n=0,1,\dots,N$,
$\alpha_n(\omega)>0$ for any $\omega\in\R_\kappa$, i.e.,
the matrix $\tilde D(\omega)$ is positive definite for $\omega\in\R_\kappa$.

(A4) For every $n=0,1,\dots,N$,
$\beta_n(\omega)>0$ for any $\omega\in\R_\kappa$.

(A5) For every $n=0,1,\dots,N$,
$c_n(\omega)>0$ for any $\omega\in\R_\kappa$.

(A6) $\alpha_n(\kappa)>0$ for $n=0,1,\dots,N-1$, and $\alpha_N(\kappa)\ge0$.

(A7) $c_n(\kappa)>0$ for $n=1,\dots,N-1$, and $c_N(\kappa)\ge0$.
\end{lemma}
{\bf Proof}\, For $\omega\in\R_\kappa$,
all functions $d_{n}(\omega)\in\R$ and $d_{n}(\omega_1)<d_{n}(\omega_2)$
for $|\omega_2|<|\omega_1|<\kappa$.
 Hence,
 all $c_n(\omega)$ are even, strictly decrease for $\omega\in[0,\kappa)$.
Furthermore, $\tilde D(0)$ is positive definite by Lemma~\ref{l5.5}.
Hence,
(A1) $\Longleftrightarrow$ (A2);
(A3) $\Longrightarrow$ (A2);
(A3) $\Longleftrightarrow$ (A4) $\Longleftrightarrow$ (A5);
(A7)~$\Longrightarrow$~(A5);
(A6) $\Longrightarrow$ (A2);
(A6) $\Longleftrightarrow$ (A7).
The proof of implications (A2) $\Longrightarrow$ (A3)
and (A2)~$\Longrightarrow$~(A6)
can be proved by a similar way as Lemma~\ref{l5.3}.
\bo
\begin{remark}
(i) If ${\rm det }\,\tilde D(\omega)\not=0$
for any $\omega\in\R: |\omega|>a$, then
$d_{n}(\omega)< 0$ for $|\omega|\ge a$.
Furthermore, $(-1)^n(\alpha_n(\omega))'<0$ for $\omega>a$.
\smallskip

(ii) If ${\rm det }\,\tilde D(\omega)\not=0$
for any $\omega\in\R: |\omega|<\kappa$, then
$d_{n}(\omega)> 0$ for $|\omega|\le\kappa$.
\smallskip

(iii)  If assertions~(A2) of Lemmas~\ref{l5.3} and \ref{l5.4} are not fulfilled, then
 there exists a point $\omega_*\in\R\setminus\{\Lambda\cup0\}$ such that
 $\textrm{det }\tilde D(\omega_*)=0$ and the entries of the inverse matrix
 $\tilde {\cal N}(\omega)$ have poles
 at points $\omega=\pm\omega_*$. Then,  bound~(\ref{NN}) does not hold.
\smallskip

(iv) By Lemmas~\ref{l5.3} and \ref{l5.4}, $\alpha_0(a)<0$ and $\alpha_0(\kappa)>0$.
Then, $m_0>m/2$.
Similarly, since $\beta_N(a)<0$ and $\beta_N(\kappa)>0$, then $m_N>m/2$.
\end{remark}

\subsection{Properties of $\tilde D(\omega)$ and $\tilde {\cal N}(\omega)$}
\label{appendixC}

Let $N\ge1$. If $\omega\in\R\setminus \Lambda$, then $\tilde D(\omega)\in\R$.
Furthermore, $\tilde D(\omega)$ is a  symmetric tridiagonal matrix
with  non-diagonal entries $\tilde D_{ii+1}(\omega)=-\gamma_i^2\not=0$.
If $\tilde D(\omega)\in\R$, then $\tilde D(\omega)$ is a well-known normal Jakobi  matrix.
Tridiagonal matrices are widely studied in the literature, see e.g., \cite{GK, Usmani}.
We mark some interesting facts on the matrix $\tilde D(\omega)$.
Write $d_i\equiv d_i(\omega)=\tilde D_{ii}(\omega)\in\R$.
To find the determinant of $\tilde D(\omega)$, we can use the three-term recurrence relation
\be\label{4.10}
\alpha_i=d_{i}\alpha_{i-1}-\gamma_{i-1}^{2}\alpha_{i-2},\quad i=0,\dots,N,
\ee
with initial conditions
$\alpha_{-2}=0$ and $\alpha_{-1}=1$. Here
 $\alpha_i$, $i=0,\dots,N$, is  the $i$-th
leading (principal corner) minor of the matrix $\tilde D(\omega)$, i.e.,
\begin{equation}\label{alphai}
\alpha_i={\rm det}\left(
  \begin{array}{ccccc}
    d_{0}&-\gamma_0&0&\ldots&0 \\
    -\gamma_0 & d_{1}& -\gamma_1& \ddots&0\\
    0&-\gamma_1&d_{2}&\ddots&0\\
    \vdots&\ddots&\ddots&\ddots&-\gamma_{i-1}\\
    0 &\ldots&0&-\gamma_{i-1}& d_{i} \\
  \end{array}
\right)
\end{equation}
Similarly, introduce the sequence $\{\beta_{i}\}_{i=0}^N$ of the determinants by the rule
\begin{equation}\label{betai}
\beta_i={\rm det}\left(
  \begin{array}{ccccc}
    d_{i}&-\gamma_i&0&\ldots&0 \\
    -\gamma_i & d_{i+1}& -\gamma_{i+1}& \ddots&0\\
    0&-\gamma_{i+1}&d_{i+2}&\ddots&0\\
    \vdots&\ddots&\ddots&\ddots&-\gamma_{N-1}\\
    0 &\ldots&0&-\gamma_{N-1}& d_{N} \\
  \end{array} \right)
\end{equation}
Then $\beta_i$ satisfies the three-term recurrence
\begin{equation}\label{4.11}
\left\{\begin{array}{ll}
\beta_i= d_{i}\beta_{i+1}-\gamma_{i}^{2}\beta_{i+2},\quad i=0,\dots,N,\\
\beta_{N+1}=1,\quad \beta_{N+2}=0.
\end{array} \right.
\end{equation}
In particular, ${\rm det}\tilde D(\omega)=\alpha_N(\omega)=\beta_0(\omega)$.
The entries of the symmetric matrix $\tilde {\cal N}(\omega)=\left(\tilde D(\omega)\right)^{-1}$
 are of a form (see \cite[Theorem~1]{Usmani})
\begin{equation}\label{Nij}
\tilde {\cal N}_{ij}(\omega)=\left\{
 \begin{array}{lll}
\gamma_{j}\gamma_{j+1}\dots\gamma_{i-1}\ds\frac{\alpha_{j-1}(\omega)\beta_{i+1}(\omega)}{\alpha_N(\omega)},
&j=0,1,2,\dots,i-1,\\
\ds\frac{\alpha_{i-1}(\omega)\beta_{i+1}(\omega)}{\alpha_N(\omega)}&j=i
 \end{array}
\right.
\end{equation}
$i=N,N-1,\dots,0$.

\begin{remark}\label{r5.6}
(1) Note that if $\alpha_n=0$ for some $n=1,2,\dots$,
then $\alpha_{n-1}\alpha_{n+1}<0$.
The last inequality can be proved using induction, recurrence (\ref{4.10})
and the fact that all $\gamma_n>0$.
\medskip

(2)
The following relation holds, $\alpha_k(\omega)=c_0(\omega)\cdot\dots\cdot c_k(\omega)$.
In particular,
\be\label{4.17}
{\rm det}\,\tilde D(\omega)\equiv \alpha_N(\omega)
=c_0(\omega)\cdot\ldots\cdot c_N(\omega),
\ee
where
\be\label{4.18}
c_n\equiv c_n(\omega)=\left\{
\ba{lll}
d_{0}(\omega),&  n=0,\\
d_{n}(\omega)-\ds\frac{\gamma^2_{n-1}}{c_{n-1}}, &n\not=0.
\ea
\right.
\ee

(3) $c_n={\alpha_n}/{\alpha_{n-1}}$, $n=0,1,\dots,N$.

(4)
The following relation holds, $\beta_k(\omega)=e_k(\omega)\cdot\dots\cdot e_N(\omega)$.
In particular,
$$
{\rm det}\,\tilde D(\omega)\equiv \beta_0(\omega)
=e_0(\omega)\cdot\ldots\cdot e_N(\omega),
$$
where
$$
e_n\equiv e_n(\omega)=\left\{
\ba{lll}
d_{N}(\omega),&  n=N,\\
d_{n}(\omega)-\ds\frac{\gamma^2_{n}}{e_{n+1}}, &n\not=N,
\ea
\right.
$$
and $e_n={\beta_n}/{\beta_{n+1}}$, $n=0,1,\dots,N$.
\end{remark}

\subsection{Asymptotics of ${\cal N}(t)$ for large times}
To prove Theorem~\ref{l3.1}, it remains to study the behavior of $\tilde {\cal N}(\omega)$
near singular points in $\Lambda^0$.
\smallskip

Let $\mu=0$. Then $\kappa=0$ and $0\in\Lambda^0$. We apply the representation (\ref{a10}) to $e^{i\theta(\omega)}$ and obtain that
$$
\tilde D_{nn}(\omega)\equiv d_n(\omega)=d_{n}(0)-i\omega\sqrt{m\gamma}
-\omega^2(m_n-m/2)+\dots\,\,\, \mbox{as }\,\,\omega\to0,\,\, \mbox{for }\,n=0,\,\,n=N,
$$
where $d_{0}(0)=\mu_0+\gamma_0$,
$d_{N}(0)=\mu_N+\gamma_{N-1}$.
Using (\ref{4.10}) and (\ref{4.11}), we find the asymptotics of determinants $\alpha_k(\omega)$ and $\beta_k(\omega)$  as
\beqn\label{alphabeta}
\left.\begin{array}{lllll}
\alpha_k(\omega) &=&\alpha_k(0)-iA_k(0)\sqrt{m\gamma}\,\omega+\dots,&k=0,\dots,N-1,  \\
\beta_k(\omega)&=& \beta_k(0)-iB_k(0)\sqrt{m\gamma}\,\omega+\dots,&k=1,\dots,N,
\end{array}
\right|\,\,\,\omega\to0,
\eeqn
where $\alpha_k(0)>0$ for $k=0,\dots,N-1$, $\beta_k(0)>0$ for $k=1,\dots,N$,
$A_0\equiv1$, $B_N\equiv 1$,
$A_k(\omega)=\Delta^1_k$,
$B_k(\omega)=\Delta^k_{N-1}$, $k=1,\dots,N-1$,
where $\Delta_k^j$ are defined in (\ref{deltai}).


If $\mu_0=\dots=\mu_N=0$, then
$\textrm{det }\tilde D(0)=0$ and
$$
\alpha_N(\omega)\equiv\textrm{det }\tilde D(\omega)=-iC_0\sqrt{m\gamma}\omega+\dots
\quad \mbox{as }\,\,\omega\to0,
$$
with
$$
C_0=\beta_1(0)+\alpha_{N-1}(0)=2\beta_1(0)=2\gamma_0\cdot\dots\cdot\gamma_{N-1}.
$$
Furthermore,
$$
\ba{ll}
\alpha_k(0)=\gamma_0\cdot\ldots\cdot\gamma_{k},\quad k=0,\dots, N-1,\\
\beta_k(0)=\gamma_{k-1}\cdot\ldots\cdot\gamma_{N-1},\quad k=1,\dots, N.
\ea
$$
Therefore, by (\ref{Nij}) and (\ref{alphabeta}),
the entries
$\tilde {\cal N}_{nk}(\omega)$ have a simple pole at zero,
\be\label{4.200}
\tilde {\cal N}_{nk}(\omega)=
\frac{i}{2\sqrt{\gamma m}\,\omega}+C_{nk}+\dots\quad \mbox{as }\,\,\omega\to0,
\quad n,k=0,\dots,N,
\ee
where  $C_{nk}$ are some constants.
If there is nonzero $\mu_n$ for some $n\in\{0,\dots,N\}$, then
$$
\textrm{det }\tilde D(0)\ge \mu_n\gamma_0\cdot\ldots\cdot\gamma_{N-1}>0.
$$
Therefore,
(\ref{lemmaA3}) follows by conditions ${\bf C}$ and ${\bf C}_0$.
\medskip

Let $\kappa\not=0$. Then
we apply the representation (\ref{a8}) to $e^{i\theta(\omega)}$ and obtain
$$
d_{n}(\omega)=d_{n}(\kappa)
-i\sqrt{m\gamma}(\omega^2-\kappa^2)^{1/2}+\dots\nonumber
\quad\mbox{as }\,\, \omega\to\kappa,\,\,\, \omega\in\mathbb{C}_+,
\,\,\, n=0;N,
$$
where
 $d_{0}(\kappa)=\mu_0-m_0\kappa^2+\gamma_0$,
  $d_{N}(\kappa)=\mu_N-m_N\kappa^2+\gamma_{N-1}$.
 Therefore, the asymptotics of determinants $\alpha_k(\omega)$ and $\beta_k(\omega)$   is
\beqn\nonumber
\left.\begin{array}{lllll}
\alpha_k(\omega) &=&\alpha_k(\kappa)-iA_k(\kappa)\sqrt{m\gamma}(\omega^2-\kappa^2)^{1/2}+\dots,&k=0,\dots,N-1,  \\
\beta_k(\omega)&=& \beta_k(\kappa)-iB_k(\kappa)\sqrt{m\gamma}(\omega^2-\kappa^2)^{1/2}+\dots,&k=1,\dots,N.
\end{array}
\right|\,\,\,\omega\to\kappa.
\eeqn
If $\textrm{det }\tilde D(\kappa)\not=0$, then condition~${\bf C}$
holds and we obtain the representation~(\ref{lemmaA1}).
If  $\textrm{det }\tilde D(\kappa)=0$, then
condition~${\bf C}_0$
holds and we obtain the representation~(\ref{lemmaA2}).
\smallskip

For $\omega\to a$, we apply the representation (\ref{a9}) to $e^{i\theta(\omega)}$ and obtain
$$
d_{n}(\omega)=d_{n}(a)-i\sqrt{m\gamma}(a^2-\omega^2)^{1/2}+\dots
\,\,\,\mbox{as }\,\, \omega\to a,\,\,\, \omega\in\mathbb{C}_+,
\,\,\, n=0;N,
$$
where
 $d_{0}(a)=\mu_0-m_0a^2+2\gamma+\gamma_0$,
 $d_{N}(a)=\mu_N-m_Na^2+2\gamma+\gamma_{N-1}$.
If $\textrm{det }\tilde D(a)\not=0$, then condition~${\bf C}$
holds and we obtain the representation~(\ref{lemmaA1}),
if  $\textrm{det }\tilde D(a)=0$, then
condition~${\bf C}_0$
holds and we obtain the representation~(\ref{lemmaA2}).
Therefore, Theorem~\ref{l3.1} in the case $N\ge1$ can be proved
by a similar way as in the case  $N=0$, see Appendix~A.

\subsection{Resonance cases:  $N\ge1$}\label{Sec4.6}
In the case of $N=0$, the resonance cases are considered in Section~\ref{Sec3.6}.
Now we consider the case $N\ge1$ and
 construct the solutions $u(n,t)$ which do not satisfy the bound (\ref{2.31'}).
 If conditions~{\bf C} and ${\bf C}_0$ are not satisfied, then
 there are two possible cases:

 \emph{(1)} $\mu=\mu_0=\dots=\mu_N=0$;

  \emph{(2)} There is a point $\omega_0\in\R\setminus\Lambda$ such that
${\rm det}\,\tilde D(\omega_0)=0$.
\smallskip

  In the case~\emph{(1)},
   $\tilde {\cal N}_{nk}(\omega)$ have a simple pole at zero and, by (\ref{4.200}),
$$
{\cal N}_{nk}(t)=(2\sqrt{\gamma\,m })^{-1}+O(t^{-1/2})\quad\mbox{ as }\,\,t\to\infty,
\quad n,k=0,1,\dots,N.
$$
Suppose that the initial data $Y_0(n)\equiv0$ for $n\not=\{0,\dots,N\}$.
Then, $Y_0\in{\cal H}_{\alpha}$ for any $\alpha$ and $z(n,t)\equiv0$ for any $n$,
where $z(n,t)$ is defined in (\ref{2.3}).
Using (\ref{2.1}) and (\ref{2.34}), we obtain
$$
u(0,t)=r(0,t)=\frac{1}{2\sqrt{\gamma\,m}}\sum\limits_{k=0}^N v_0(k)
+O(t^{-1/2})\quad\mbox{as }\,\, t\to\infty.
$$
Hence, if $v_0(0)+\dots +v_0(N)\not=0$, then
the solution $u(\cdot,t)$ does not satisfy the bound~(\ref{2.31'}).
\medskip

 In  the case \emph{(2)}, there is a nonzero vector $\xi=(\xi_0,\dots,\xi_N)\in\R^{N+1}$
such that ${\rm det}\,\tilde D(\omega_0)\xi=0$.
Note that for $\omega_0\in\R\setminus \Lambda$,
$\Im\theta(\omega_0)>0$, $\Re\theta(\omega_0)=\pm\pi$.
Therefore, the function of the form $u(n,t)=v(n)\sin(\omega_0 t)$, where
$$
v(n)=\left\{
\ba{ll}
e^{-i\theta(\omega_0)n}\,\xi_0&\mbox{if }\,\,n\le0 \\
\xi_n   &\mbox{if }\,\,n=1,\dots,N-1 \\
e^{i\theta(\omega_0)(n-N)}\,\xi_N &\mbox{if }\,\,n\ge N,
\ea\right.
$$
is a solution of the system
with the initial data  $Y_0=(0,v_0)$,
where $v_0(n)=\omega_0 m_n v(n)$ with $m_n=m$ for $n\le -1$ and $n\ge N+1$.
 Note that $Y_0\in{\cal H}_{\alpha}$ for any $\alpha$.
However,
$\Vert u(\cdot,t)\Vert_{\alpha}=\Vert v\Vert_{\alpha} |\sin(\omega_0 t)|\ge C|\sin(\omega_0 t)|$
with some constant $C>0$.
Therefore, the bounds~(\ref{0.3}) and (\ref{2.31'}) are not fulfilled for this solution.



\end{document}